\documentclass{optica-article}

\journal{opticajournal} 

\articletype{Research Article}

\usepackage{amsmath}
\usepackage{graphicx}
\usepackage{subcaption}
\usepackage{caption}
\DeclareMathOperator{\sinc}{sinc}
\usepackage{float}
\usepackage{sidecap}
\usepackage{siunitx} 
\usepackage{hyperref}
\usepackage{multirow}
\usepackage{amsmath}
\usepackage{enumitem}
\usepackage{booktabs}
\usepackage{setspace}
\newcommand{\insertfigure}[3]{\begin{subfigure}{#1\textwidth}
		\centering
		\includegraphics[width=\textwidth]{#2}
		\caption{ } 
		\label{#3}
\end{subfigure}}

\newcommand{\NA}{\mathit{N\hspace{-0.2em}A}}  
\newcommand{\lem}{\lambda_{\text{em}}} 
\newcommand{\hamp}{\vec{h}_{amp}} 

\newcommand{\todo}[1]{}
\renewcommand{\todo}[1]{{\color{red} TODO: {#1}}}
\newcommand{\rh}[1]{}
\renewcommand{\rh}[1]{{\color{gree} Rainer: {#1}}}

\usepackage{algorithm}
\usepackage{algpseudocode}
\usepackage{algorithmicx}
\algdef{SE}[DOWHILE]{Do}{doWhile}{\algorithmicdo}[1]{\algorithmicwhile\ #1}%

\algrenewcommand\algorithmicrequire{\textbf{Input:}}
\algrenewcommand\algorithmicensure{\textbf{Output:}}

\begin{document}

\title{Calculating point spread functions: methods, pitfalls and solutions}

\author{Ratsimandresy Holinirina Dina Miora,\authormark{1,2} Erich Rohwer,\authormark{1} Martin Kielhorn,\authormark{3} Colin Sheppard,\authormark{4} Gurthwin Bosman\authormark{1,*} and Rainer Heintzmann,\authormark{2,3,5}}

\address{\authormark{1} Physics Department, Stellenbosch University, Private Bag X1, Matieland 7602, South Africa \\
\authormark{2} Faculty of Physics and Astronomy, Friedrich-Schiller-Universit\"at Jena, Max-Wien-Platz 1, 07743 Jena, Germany \\
\authormark{3} Institute of Physical Chemistry and Abbe Center of Photonics, Friedrich-Schiller-Universit\"at Jena, Helmholtzweg 4, 07743 Jena, Germany\\
\authormark{4} Nanoscopy \& NIC@IIT, Istituto Italiano di Tecnologia, Via Enrico Melen, 83 Edificio B, 16152 Genova, Italy\\
\authormark{5}Leibniz Institute of Photonic Technology, Albert-Einstein Str. 9, 07745 Jena, Germany}

\email{\authormark{*}heintzmann@gmail.com}

\begin{abstract*} 
The knowledge of the exact structure of the optical system PSF enables a high-quality image reconstruction in fluorescence microscopy. Accurate PSF models account for the vector nature of light and the phase and amplitude modifications. Most existing real-space-based PSF models fall into a sampling pitfall near the centre position, yielding to the violation of the energy conservation. In this work, we present novel, to the best of our knowledge, Fourier-based techniques for computing vector PSF and compare them to the state-of-the-art. Our methods are shown to satisfy the physical condition of the imaging process. They are reproducible, computationally efficient, and easy to implement and easy to modify to represent various imaging modalities.
\end{abstract*}

\section{Introduction}\label{intro}


Many studies have been conducted to represent, model and approximate the impulse response of a fluorescent microscope, the point spread function (PSF) \cite{gibson1989diffraction, li2017fast, RW1, RW2, kirshner20133, aguet2009super, samuylov2019modeling, torok1997electromagnetic}. An accurate PSF model is an important requirement for a successful image reconstruction \cite{sage2017deconvolutionlab2}. A PSF model can be based on scalar diffraction theory or account for the vectorial nature  of the electromagnetic field, such as the polarization state of the field. The former generally outperforms the latter in terms of computation time, but its validity is limited to the case of low numerical aperture (NA) optical systems, \textit{i.e.} a small maximum half opening angle, $\theta_\text{max}$, which corresponds to $\sin(\theta_\text{max}) \lesssim 0.4$   \cite{gu2000advanced}. A very popular scalar PSF model was developed by Gibson and Lanni (GL) \cite{gibson1989diffraction, gibson1991experimental}. Li \textit{et al.} implemented this scalar PSF model computationally by approximating Kirchhoff's integral using a Bessel series \cite{li2017fast}. The method is fast as it assumes radial symmetry in the PSF, however this assumption limits the generality of the type of optical aberrations that can be included in the model.

In high NA systems, $\sin(\theta_\text{max}) > 0.7$   \cite{gu2000advanced}, the vector nature of light is required to produce a realistic PSF \cite{sheppard1993imaging, RW1, RW2}. For this it is common to use the Richards and Wolf (RW) model. The RW model mathematically formalizes the most accurate approximation of the exact representation of the vector diffraction pattern in optical systems \cite{RW1, RW2}. The three integrals in its formulation are however expensive to compute. The model was modified by T\"or\"ok and Varga (TV) to include planar interfaces and refractive index mismatches \cite{torok1995electromagnetic}. The TV model was later generalized to account for a stratified medium \cite{torok1997electromagnetic}. Additionally, Haeberl\'e combined the TV formulation with the GL model to compute the PSF of a conventional microscope \cite{haeberle2003focusing}. Ghosh and Preza adapted the method developed by Haeberl\'e to account for aberrations due to spatially variant refractive indices in the sample and validated the model experimentally \cite{ghosh2015fluorescence}. The consideration assumes patches (tiles) of the sample for which the assumption of shift-invariant imaging holds.  
The RW model was also computed by Kirshner et al. \cite{kirshner20133}. A toolbox that allows free access to this model is freely available online under the name PSFGenerator \cite{kirshner20113d, psfgensoftware}. 
This toolbox has been widely used for deconvolution \cite{sage2017deconvolutionlab2, guo2020rapid}. However, based on our experience, the software does not support large window sizes and the computation is not time- and memory efficient. Compared to the other PSF models, we also observed that the violation of the missing cone near the focus is more pronounced in PSFs computed using this toolbox. Aguet et al. also presented a method for PSF computation \cite{aguet2009super}. They use a numerical integration based on Simpson's rule to compute the scalar and vector PSF. A recent work in PSF modeling consists of computing the PSF by combining several weighted Gaussian kernels shifted to different positions \cite{samuylov2019modeling}. The efficiency and accuracy of this model critically depends on the number of  single Gaussian functions used. Lastly, a free and open-source Python software package for PSF calculation, PyFocus, was also released \cite{caprile2022pyfocus}. This software calculates the vector field beyond the paraxial approximation using a custom-made 2D trapezoidal integration algorithm.    

Despite the numerous advantages of many of the PSF models described above, each model still has its limitations. 
Many models exploit radial symmetry to achieve fast computation. Yet, such models cannot directly support asymmetric aberration that may be present in the system. Non-radially symmetric aberrations often include coma and astigmatism. In this work, we present four Fourier-based, comprehensive, easy to implement and more realistic methods for computing PSFs, two of which are completely new to the literature to the best of our knowledge. The use of Fast Fourier Transform (FFT) on a regular grid allows us to efficiently include non-radially symmetric effects. Nevertheless, the FFTs have their drawbacks (\textit{e.g.}  border wrap-around, a form of aliasing) which must be carefully accounted for. We therefore present the pitfalls that one may encounter in the calculation and ways around them. Our models also account for losses by reflection on non-coated optical surfaces (\textit{e.g.} the coverslip) via Fresnel coefficients. Additionally, the Fourier plane can be easily accessed and sample-induced aberrations that may arise during imaging can also be introduced into the PSF calculation. The models are demonstrated to be fast compared to state-of-the-art models supporting similar features. They can easily be modified to represent different imaging modalities or to include a tilt of optical surfaces in the imaging system. Common situations such as a tilt of the coverslip can thus be modelled. A further aim of this manuscript is to release a toolbox to the scientific community, which others can benefit from for calculating PSFs using uniform or modified (aberrant) apertures \cite{PSFToolbox}. The software code is written with MATLAB programming language and is released in an open source format. It can easily be converted to any other programming language such as Python.

Before we introduce the four Fourier-based models in Section \ref{methods}, we describe the necessary tools that are needed for their computation in Section \ref{tools}. We then conduct a quantitative comparison of the models with the state-of-the-art in Section \ref{comparison}. We conclude our results in Section \ref{conclusion}.


\section{Tools for computing a vector PSF using Fourier optics}\label{tools}
\subsection{Abbe sine condition}

\begin{figure}[htbp]
	\centering
	\includegraphics[width=0.7\textwidth]{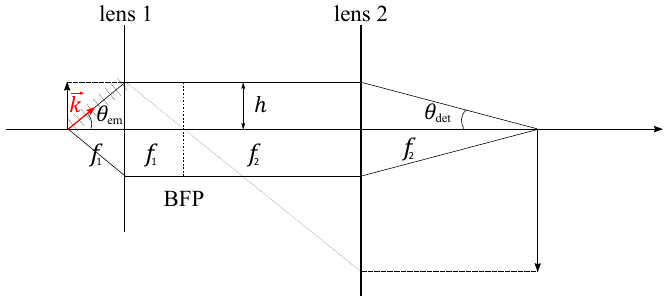}
	\caption{Geometrical representation of a $4$f-imaging system using the thin lens approximation. Note that here $\tan(\theta_{\text{em}}) /\tan(\theta_{\text{det}}) = f_2/f_1$, which violates Abbe\textquotesingle s sine condition. BFP stands for back focal plane. Here lens 1 corresponds to the microscope objective lens while lens 2 represents the tube lens.}
	\label{fig:4f-imaging}
\end{figure}

An optical imaging system usually fulfills the Abbe sine condition to image a plane in the object to the detector plane (e.g. a CCD or CMOS camera) \cite{gu2000advanced, malacara2017opticaldesign}. The Abbe sine condition relates any angle of a beam emitted by the sample ($\theta_{\text{em}}$) to its corresponding angle reaching the detector ($\theta_{\text{det}}$). It is given by $\sin(\theta_{\text{em}}) /\sin(\theta_{\text{det}}) = M$, with $M$ being the magnification of the optical system in terms of geometric optics \cite{gu2000advanced,singer2006handbook}. As shown in the thin lens approximation in Fig. \ref{fig:4f-imaging}, a simple $4f$-imaging microscope imaging system with a magnification of $M = 2$ would not fulfill the above-mentioned Abbe sine condition. 

\begin{figure*}[htbp]
	\centering
	\includegraphics[width=0.83\textwidth]{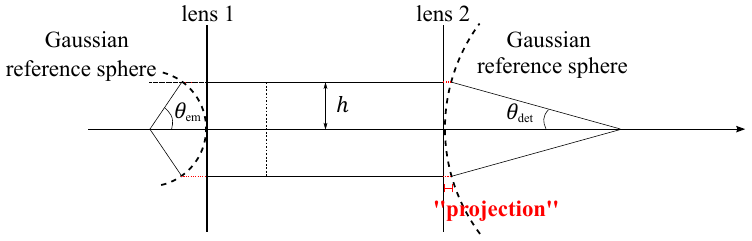}
	\caption{Representation of the Gaussian reference sphere, forcing $\sin(\theta_{\text{em}})/\sin(\theta_{\text{det}})=f_2/f_1$ to be constant in agreement with Abbe\textquotesingle s sine condition. "Teleportation" means that the beams are continued at the connected surface without acquiring any phase for the space in between the plane normal to the optical axis and the Gaussian reference sphere.}
	\label{fig:4f-imagingGaussian}
\end{figure*}

The concept of Gaussian reference sphere, defined as the collection of equivalent refractive loci for an aplanatic imaging system, is introduced to maintain the simplicity of such drawings (see Fig. \ref{fig:4f-imagingGaussian}). This sphere is centred at the intersection point of the object plane or the image plane with the optical axis. Its radius is equal to the focal length $f_1$ or $f_2$ of the respective imaging lenses \cite{singer2006handbook}. The rays propagate to the Gaussian reference sphere and get projected to a plane normal to the optical axis without acquiring any extra phase. This projection is indicated by the red dashed lines in Fig. \ref{fig:4f-imagingGaussian}. The same principle is applied in reverse to the tube lens (lens 2 in Fig. \ref{fig:4f-imagingGaussian}), where the effect can often be neglected due to the usually large imaging magnification and thus small angles near the image plane (see right side of Fig. \ref{fig:4f-imagingGaussian}).

\subsection{Energy conservation}

The energy of light propagating through the aplanatic focusing system must be conserved. Integrating over the radial position in the back focal plane (see Fig. \ref{fig:4f-imaging} and \ref{fig:4f-imagingGaussian}) must yield to the same quantity of energy as integrating over the corresponding angle $\theta$ in the equivalent refractive locus. Given the projection of the field in the Gaussian reference sphere onto the normal planes, the aplanatic factor (AF) needs to be accounted for to conserve the energy. 

\begin{figure*}[htbp]
	\centering
	\includegraphics[width=0.5\textwidth]{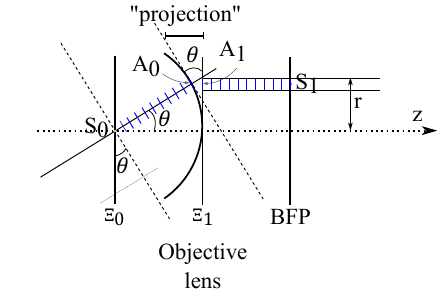}
	\caption{Schematic diagram illustrating the ``aplanatism effect''. The two mutually parallel planes $\Xi_0$ and $\Xi_1$ are perpendicular to the optical axis. The area $A_0$ is a projection of the elementary area $A_1$ of the plane $\Xi_1$ onto the Gaussian reference sphere.}
	\label{fig:aplanatic factor}
\end{figure*}

To understand and derive the AF, let us assume an isotropic emitter placed at the centre of the Gaussian reference sphere, $S_0$ (see Fig. \ref{fig:aplanatic factor}). Since the emitter is emitting uniformly in all directions, the strength of the amplitudes on the Gaussian reference sphere is uniform. Let us also consider a small parallel ray which is redirected from the Gaussian reference sphere at a given incidence angle $\theta$. If we attribute a given irradiance $I_0$ as a power $P_0$ per unit area $A_0$ to such a beamlet from $S_0$, the same power will have to be distributed to a smaller area  $A_1=A_0\cos(\theta)$ after the projection of the beam from the Gaussian reference sphere to the plane parallel to the pupil plane in the BFP  (see Fig. \ref{fig:aplanatic factor}). This means that the irradiance measured perpendicular to the local direction of propagation changes to $I_0/\cos(\theta)$, and each of the field vector component for this beamlet thus needs to change by $1/\sqrt{\cos(\theta)}$ to be consistent with this intensity change.  

If a large enough magnification of the imaging system is assumed, the vectorial and aplanatic factor effects caused by the tube lens can be neglected. 

To understand the factor that needs to be applied to conserve the energy when focussing a uniformly illuminated 2D pupil (BFP) with a high-NA objective (excitation PSF), let us consider the same Fig. \ref{fig:aplanatic factor} but with the light travelling from the right side to the left. The point $S_1$ denotes a point in the pupil plane from which a beamlet emerges. The parallel beamlet will focus onto $S_0$ after the objective lens. We can consider the Helmholtz reciprocity theorem: a lossless (non-magnetic) monochromatic optical system in which a field $(E_{x_0},E_{y_0},E_{z_0})$ (here an isotropic emitter) at object plane position $S_0$ gives rise to a field $(E_{x_1},E_{y_1},E_{z_1})$ at a point in the image plane via a virtual aperture around $S_1$, warrants that placing the isotropic emitter $(E_{x_0},E_{y_0},E_{z_0})$ as a source at that same point in the image plane, generates the field $(E_{x_1},E_{y_1},E_{z_1})$ at the object plane position $S_0$ \cite{sheppard1993imaging} with the same virtual aperture at $S_1$. A uniform emitter near the focus of the tube lens, leads, due to the low NA of the tube lens to a uniform illumination of the pupil plane (BFP), corresponding to our assumption for calculating the excitation PSF. Therefore the excitation PSF should have the same aplanatic correction as the emission PSF as long as the magnification is large so that we can neglect the aplanatic factor of the tube lens.

This is confirmed also by considering the $\sqrt{\cos\theta}$ dependence on the Gaussian reference sphere and thus on the McCutchen pupil (detailed in the next section). To arrive at the 2D Fourier-transform of the in-focus excitation field, we need to project the McCutchen pupil along $k_z$ and thus apply a projection factor of $1/cos\theta$ which leads to an overall amplitude of $\sqrt{\cos\theta}/\cos\theta = 1/\sqrt{\cos\theta}$ confirming the above reciprocity argument.

\subsection{The McCutchen pupil}\label{McCutchen pupil}

\begin{figure}[h]
	\centering
	\includegraphics[width=0.42\textwidth]{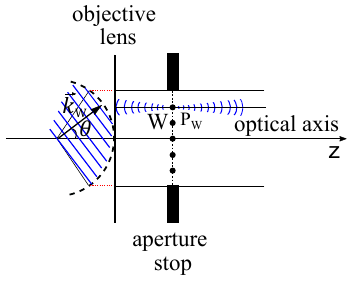}

	\caption{Visualization of the pupils and representation of the Huygens Wavelet becoming a plane wave.}
	
	\label{fig:HuygensWavelet}
\end{figure}

To calculate the PSF using a Fourier optics, let us consider the diagram in Fig. \ref{fig:HuygensWavelet} depicting a Huygens wavelet contributing to the focusing of a monochromatic coherent plane wave by a high-NA microscopy objective. The beam enters the objective system from the right side and is spatially limited by the entrance pupil of the optical system. An ``aperture stop'', is in practice either intentionally introduced to warrant the linear shift invariant performance of the system and to avoid aberrations from unwanted beams, or effectively provided by the inner geometry of the objective. The limitation of the beams is therefore approximated to be at the limit of that aperture stop. At this pupil plane, every point $P_W$ on the wavefront is considered as a source of a Huygens wavelet, denoted by W \cite{crew2009wave}. Each such Huygens wavelet is approximated by a plane wave in object space, as seen by the phase fronts within the Gaussian reference sphere. For the process of emission, which we now consider, we can interpret Fig. \ref{fig:HuygensWavelet} by decomposing the emitted wave into the same plane waves in object space, corresponding to pupil-plane wavelets.

At first, we limit ourselves to a scalar electric field, where the field is described by a single amplitude value as a function of spatial position. The vector nature of the electric field will be considered further down below. The aperture plane can be seen as a coherent superposition of spherical wavelets, each of which gives rise to a plane wave in object space directed towards the nominal focus point S (see the wavelet labelled ``W'' in Fig. \ref{fig:HuygensWavelet}). According to the Huygens-Fresnel principle, the spherically converging wave is obtained by superimposing all these plane waves \cite{crew2009wave}. These superimposed wavelets have to acquire exactly the same optical path length and constructively interfere at S. In other words, the phase at the nominal focus is identical for all such wavelets and can thus be set to zero in our simulation. For convenience, we choose S as the centre of our real-space coordinate system. The wavelet W giving rise to a plane wave in focus (Fig. \ref{fig:HuygensWavelet}) can now conveniently be described in Fourier space as a single point P$_\text{W}$ (Fig. \ref{fig:HuygensWavelet}), \textit{i.e.}  a single $3$-dimensional vector ($\vec{k}_W$) in Fourier space. All such vectors necessarily reside on a sphere of a radius $k_0=2\pi/\lambda_\text{em}, \lambda_\text{em} = \lambda/n,$ $\lambda_\text{em}$ and  $\lambda$ being the vacuum wavelength corresponding to the emitted wave respectively, and $n$ the refractive index of the embedding medium. 

\begin{figure*}[htbp]
	\centering
	\includegraphics[width=0.7\textwidth]{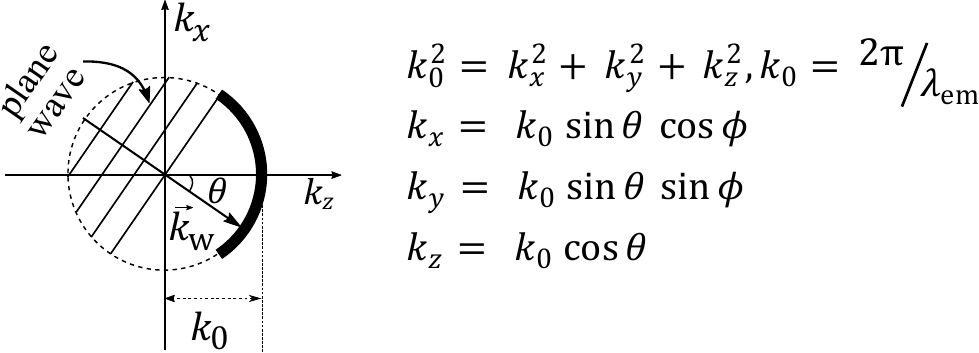}	
	\caption{Fourier space representation of a plane wave from a wavelet W at a point source P$_\text{W}$ in the pupil plane called the McCutchen pupil (bold black line). The $k_y-$axis is oriented towards the front plane of the paper. The corresponding real-space phase fronts labelled "plane wave" are shown in the Fourier-space representation to give an indicate of the direction of the wave represented by $k_W$.
     }
    	
	\label{fig:Representation of Wavelet Fourier Domain}
\end{figure*}

A pupil position in real space corresponds to the lateral components $\vec{k}_{x,y}$  of the wave-vector $\vec{k}$. This linear correspondence $\vec{p}_{x,y} = f \vec{k}_{x,y}/k_{0}$ with the focal length $f$ of the objective is forced by the Abbe sine condition between the pupil plane coordinate and the $k$-vector position of the wave near the focal plane. 

The pupil plane aperture stop thus gives rise to a three-dimensional cap residing on the $k$-sphere in Fourier space (see a section though the cap as the bold curve in Fig. \ref{fig:Representation of Wavelet Fourier Domain} in the Fourier space representation). As the aperture is limited by the NA of the objective, the $3$D frequency spectrum in Fourier space is represented in a segment of the $k$-sphere sphere. This segment is called ``generalized aperture'' or ``McCutchen pupil'' \cite{mccutchen1964generalized}.

According to McCutchen, the (complex-valued) amplitude distribution of the electromagnetic field in real space near the focus S corresponds to the three-dimensional Fourier transformation of the amplitude distribution on the McCutchen pupil \cite{mccutchen1964generalized}. To compute the intensity PSF, we can thus project the amplitude on the McCutchen pupil and perform an inverse three-dimensional Fourier transformation. The Fourier-based PSF models that are presented in this work are based on this basic understanding. The four different methods differ in how such a distribution is obtained and how the field is propagated in the homogeneous medium.

\subsection{Sampling condition of the computation}\label{sampling}

A PSF calculation typically samples the continuous mathematical function at discrete locations (delta-shaped points). The imaging of a point emitter, which is practically detected on a pixelated imaging device such as a CCD or CMOS camera. Those devices integrate, in each of their rectilinearly spaced pixels, over the continuous signal weighted by a (box-shaped) pixel sensitivity function. The local integration of the PSF in every pixel by the camera can be rewritten as first convolving the PSF with the pixel sensitivity function and then sampling it at regularly spaced positions. Due to the convolution theorem, the effect of detector integration can be represented by a multiplication of the Fourier transform of the PSF, the optical transfer function (OTF), with the Fourier transform of the pixel sensitivity function. If we assume box-pixels with uniform sensitivity, the OTF gets modified by a multiplication with a $\sinc\left(\pi k_x/k_{\text{samp}}\right)\sinc\left(\pi k_y/k_{\text{samp}}\right)$. 
This means that at the current sampling frequency $k_{\text{samp}}=2\pi/d_{\text{samp}}$, $d_{\text{samp}}$  being the pixel pitch, the overall transfer would cross zero. 
To avoid aliasing, the sampling of the PSF $d_{\text{samp}}$ must satisfy the Nyquist Shannon theorem given by: $d_{\text{samp}}<2\pi/(2k_{\text{limit}})$, with the limit frequency $k_{\text{limit}}$ referring to the transferred frequency limit, to avoid potential aliasing of information within the frequency band\cite{heintzmann2006band}.

In a wide-field system, the maximal in-plane spatial frequency is derived from Abbe diffraction limit and is given by $k_{xy,\text{max}}=4\pi\text{NA}/\lambda_{\text{em}}$  \cite{heintzmann2006band}. Similarly, the maximal axial frequency in real space for a wide-field microscope is given by $k_{z,\text{max}}=(2\pi n(1-\cos(\theta)))/\lambda_{\text{em}}$. The highest frequency of the intensity result has to be sampled with at least two positions per shortest period that can be transmitted by the system \cite{heintzmann2006band}. This requires the pupil to fit into half the digital Fourier-space representation such that its autocorrelation (\textit{i.e.} the incoherent OTF) fits in the digital Fourier space. The maximal pupil radius in Fourier space along $x$ or $y$ should be lower than half the maximally represented frequency along $k_x$ or $k_y$ in our Fourier-space representation.

\subsection{Digital Fourier transform and its pitfalls}
\subsubsection{Jinc aperture}\label{jinc-trick}

In addition to the potential error with the aforementioned sampling, a digitization of the usually round pupil in Fourier space as a hard aperture onto a rectilinear grid may also induce severe artefacts that we need to avoid when using the FFT operator. To illustrate this problem, let us consider a field distribution of a high-NA PSF with numerical aperture \mbox{NA = $1.4$}, refractive index $n = 1.518$ and emission wavelength $\lambda_{\text{em}}=580$ \si{nm}. We calculate the pupil radius, which also corresponds to the Nyquist frequency, using the theory stated in Section \ref{sampling}. We denote $r_{\text{max}}$ this pupil radius. We generate a hard aperture with radius equal to $r_{\text{max}}/8$ and calculate the corresponding field distribution in real space by generating the Fourier transform of the hard aperture. The window size for this first experiment is $128\times 128$ pixels.  

\begin{figure*}[htbp]
	\centering
	\begin{subfigure}[h]{0.24\textwidth}
		\centering
		\includegraphics[width=\textwidth]{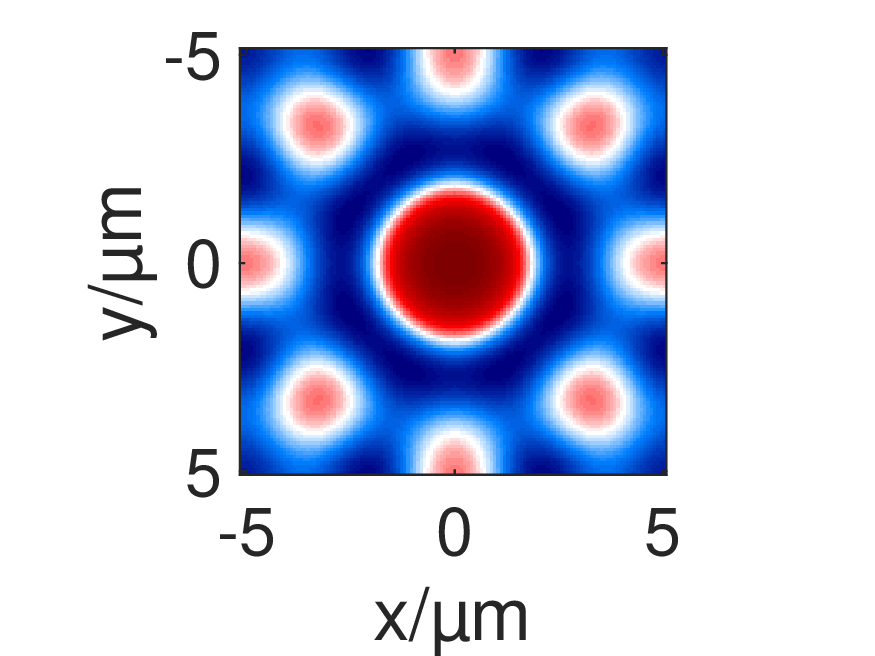}
		\caption{ }
		\label{fig:hard aperture 128}
	\end{subfigure} 
	\begin{subfigure}[h]{0.24\textwidth}
		\centering
		\includegraphics[width=\textwidth]{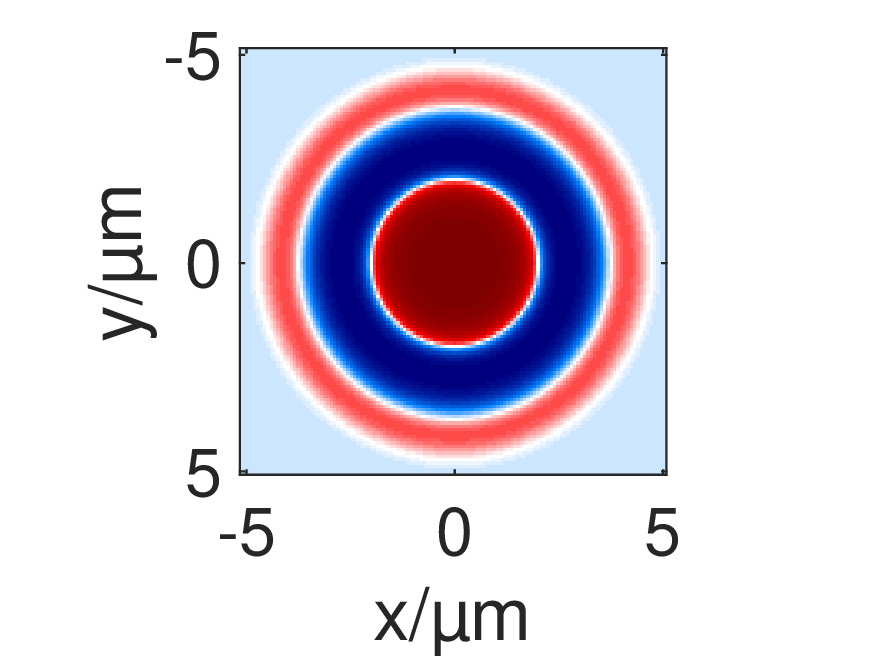}
		\caption{ }
		\label{fig:jinc aperture 128}
	\end{subfigure} 
	\begin{subfigure}[h]{0.24\textwidth}
		\includegraphics[width=\textwidth]{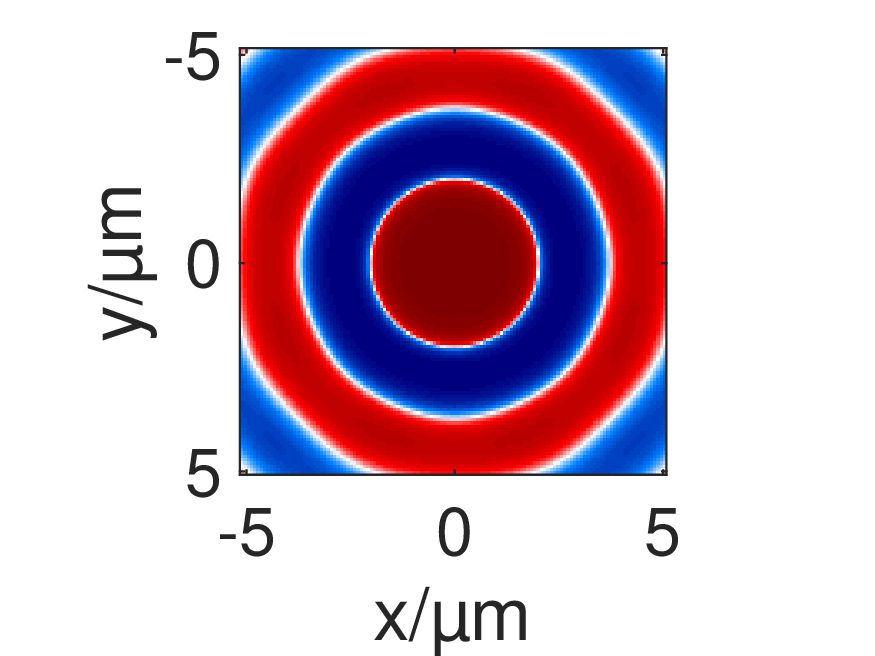}
		\caption{ }
		\label{fig:hard aperture 1024}
	\end{subfigure} 
	\begin{subfigure}[h]{0.24\textwidth}
		\centering
		\includegraphics[width=\textwidth]{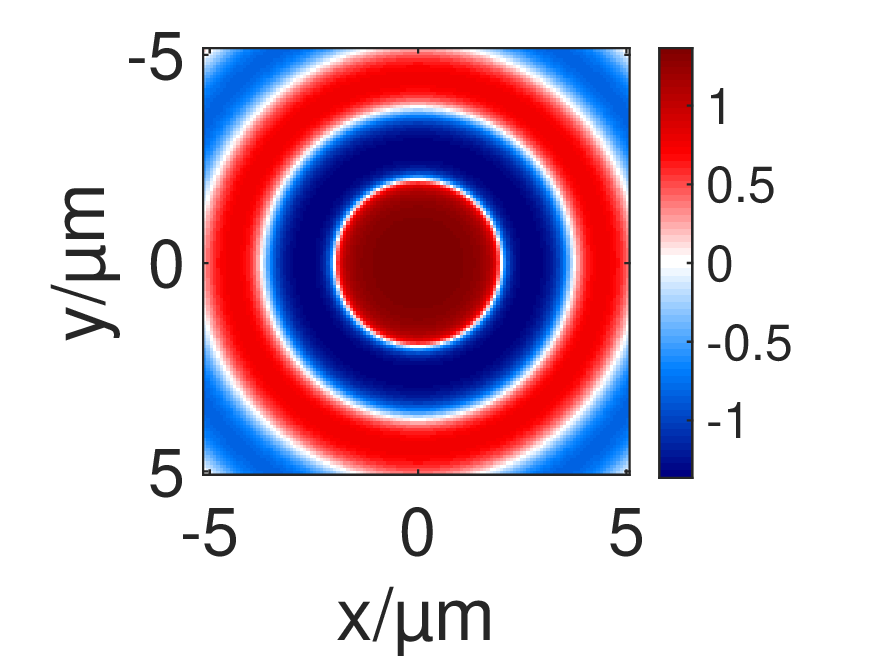}
		\caption{ }
		\label{fig:jinc aperture 1024}
	\end{subfigure} 
	\caption{Field distribution calculated from the Fourier transform of (a) a hard aperture of size $128\times 128$ pixels, (b) a jinc aperture aperture of size  $128\times 128$ pixels, (c) a hard aperture of size $1024\times 1024$ pixels cropped to $128\times 128$ pixels size for display and, (d) a jinc aperture aperture of size  $1024\times 1024$ pixels cropped to $128\times 128$ pixels size for display. A DampEdge of \SI{15}{\percent} is applied to the generated field (full size) using the jinc-trick in (b) and (d). Figures are displayed at $\tan^{-1}(\gamma E)$ and centered at the zero of the display, $E$ being the field distribution and $\gamma = 20$.}
	\label{fig:Aperture effect}
\end{figure*}

For symmetry reasons, we would expect a perfect circularly symmetric PSF. A significant deviation from circular symmetry is however clearly visible in Fig. \ref{fig:hard aperture 128}. By repeating the same calculation for $1024\times 1024$ pixels, the discrepancy is significantly reduced even though it is still not totally spherically symmetric (see Fig. \ref{fig:hard aperture 1024}). However, calculating on such large grids causes about significant computational time overhead ($0.07$ \si{s} vs $0.025$ \si{s} \textit{i.e.} about $3$ times slower). This computation might also be unnecessary as we may not need so many pixels of the PSF far away from its centre. 

Interpolation in Fourier-space using a $\sinc$ function to obtain a better representation of the pupil may be one route to re-establish spherical symmetry. However, this is a tricky business \cite{yaroslavsky1997efficient}. Here we present a different approach. We calculate the two-dimensional ($2$D) Fourier-transform of the uniform pupil disk, for which the analytical solution in real space is well known: $\text{jinc}(r)=J_1(r)/r, J_1$ being the Bessel function of the first kind and $r$ the radial coordinate. We therefore obtain an ``ideally'' representation of a disk in Fourier-space by Fourier-transforming a two-dimensional jinc function. This ``interpolated'' disk is then appropriately modified with $k$-space dependent phase and magnitude alterations. 

The computation time of the $1024\times 1024$ pixels ideal representation of disk in Fourier-space using the jinc trick is $0.3$ \si{s} on average (Windows 10, 64-bit, 
Intel\textsuperscript{\tiny\textregistered} Core\textsuperscript{\tiny TM} i5-3570S CPU @ 3.10GHz, 8,0GB RAM, Intel HD Graphics).

As the jinc-function exhibits first order discontinuities in real space at the border, causing unwanted high-frequencies \cite{cao2003generalized} in its Fourier-transform. To reduce this effect, we modify the jinc-function at the outer rim by appropriately smoothing the \SI{15}{\percent} of its edges towards zero (``DampEdge'' function in the PSFToolbox \cite{PSFToolbox}).

As seen in Fig. \ref{fig:jinc aperture 128}, the real-space representation of the field distribution is perfectly symmetric and spherical by design even for images with relatively few pixels. The circular edge of this field in Fig. \ref{fig:jinc aperture 128} has been dimmed down using the DampEdge function. We refer to this method of generating an interpolated disc in Fourier space as the jinc-FT trick. 

\subsubsection{Fourier wrap-around}\label{Fourier wrap-around}

The grid on which the field propagation is simulated is finite. There exists an axial depth position at which the disk of defocus no longer stays well within the available lateral space provided by the real-space grid. Upon propagation, waves leave on one side of the lateral sampling grid and enter on the opposite side due to the periodic boundary conditions of the Fourier-transform. This causes severe standing-wave effects called Fourier wrap-around (see Fig. \ref{fig:sp xy} and \ref{fig:sp xz}). 

\newcommand\scaleim{0.32}
\begin{figure}[htbp]
	\centering
	\begin{subfigure}[h]{\scaleim\textwidth}
		\centering
		\includegraphics[width=\textwidth]{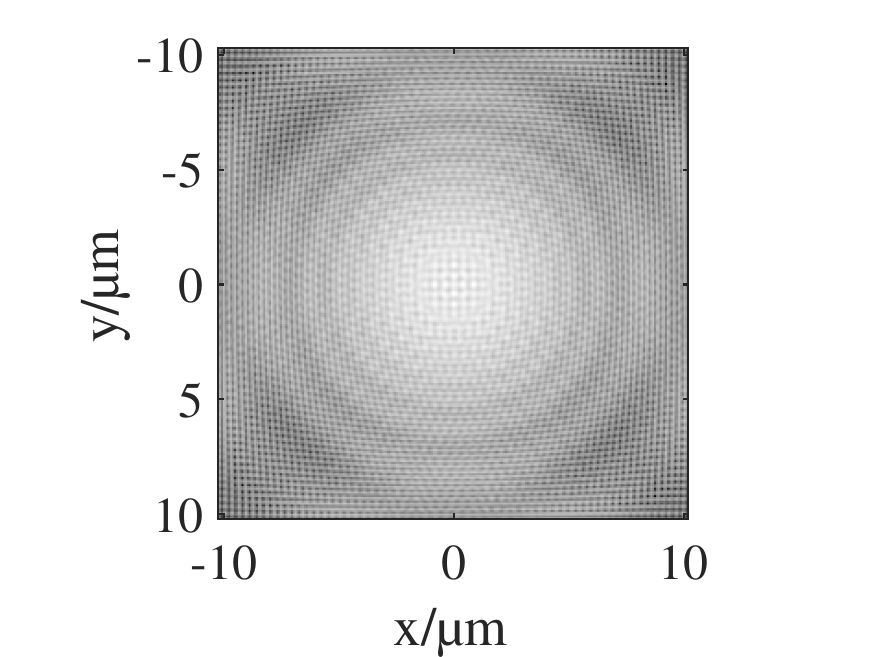}
		\caption{}
		\label{fig:sp xy}
	\end{subfigure}
	\begin{subfigure}[h]{\scaleim\textwidth}
		\centering
		\includegraphics[width=\textwidth]{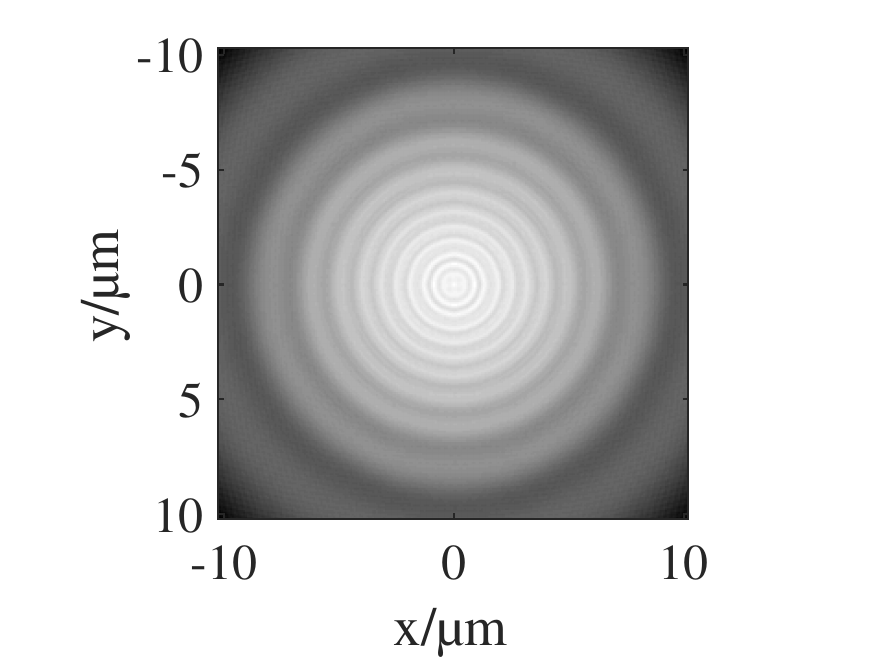}
		\caption{ }
		\label{fig:zeropad xy}
	\end{subfigure}
	\begin{subfigure}[h]{\scaleim\textwidth}
		\centering
		\includegraphics[width=\textwidth]{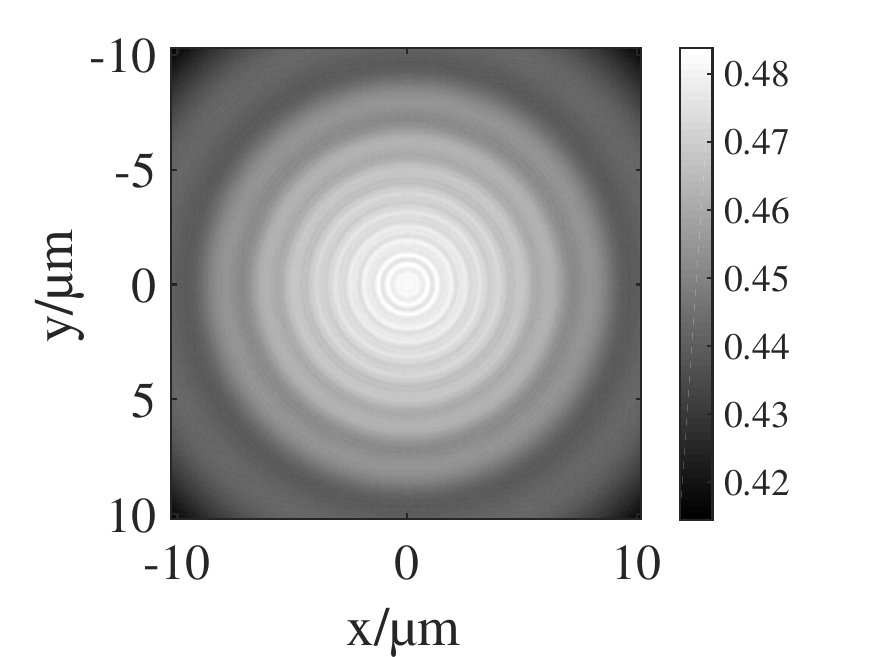}
		\caption{ }
		\label{fig:czt xy}
	\end{subfigure}
	\begin{subfigure}[h]{\scaleim\textwidth}
		\centering
		\includegraphics[width=\textwidth]{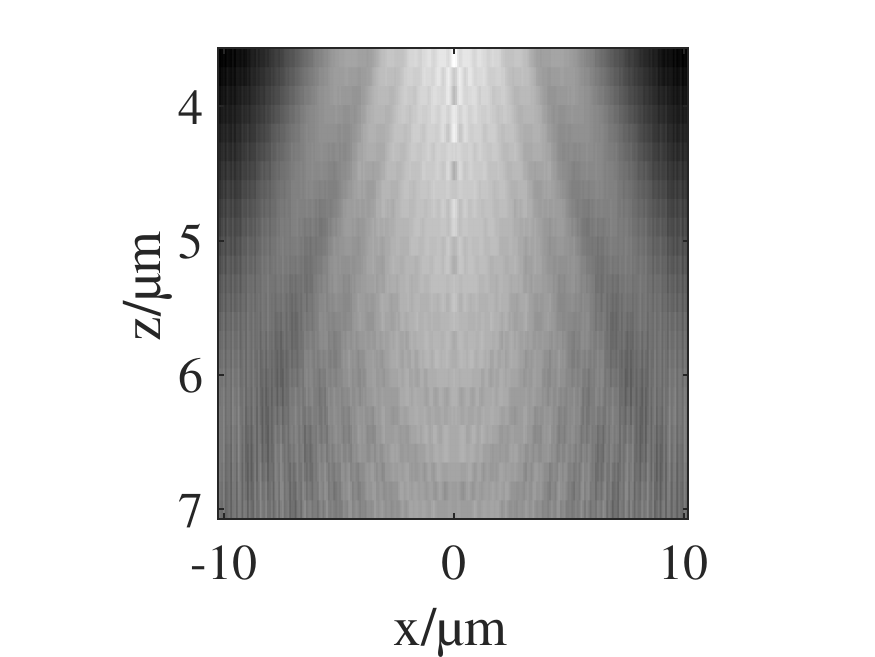}
		\caption{}
		\label{fig:sp xz}
	\end{subfigure} 
	\begin{subfigure}[h]{\scaleim\textwidth}
		\centering
		\includegraphics[width=\textwidth]{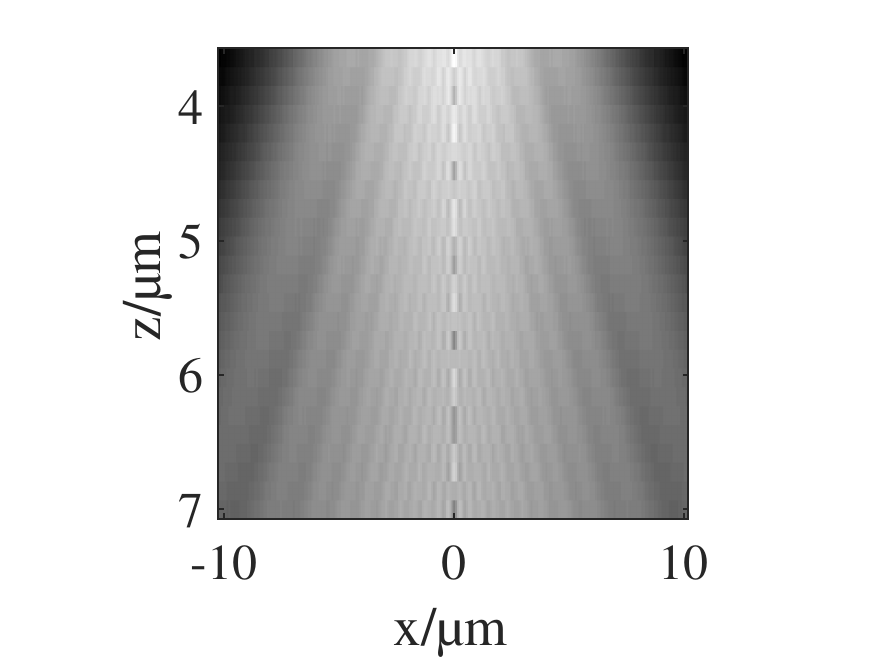}
		\caption{ }
		\label{fig:zeropad xz}
	\end{subfigure}
	\begin{subfigure}[h]{\scaleim\textwidth}
		\centering
		\includegraphics[width=\textwidth]{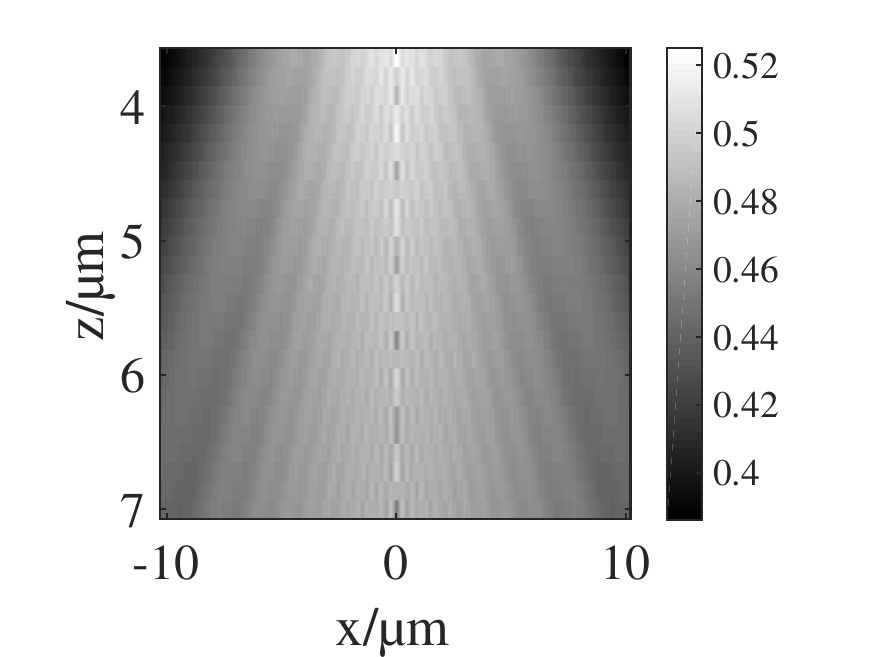}
		\caption{ }
		\label{fig:czt xz}
	\end{subfigure}
	
	\caption[Wrap-around effect using FFT-based propagation in PSF calculations]{Wrap-around of using FFTs in PSF calculations. Profiles displayed at $\gamma = 0.05$ of the PSFs calculated using the slice propagation method. (a, b, c) $xy$-plane at defocus position \SI{7}{\micro\meter}. (d, e, f) $xz$-cut parallel to the optical axis at \SI{3.5}{\micro\meter} away from the focus and with depth range $\Delta z = $ \SI{3.5}{\micro\meter}. (a, d) Directly applying the propagator in Fourier space. (b, e) By zero-padding the image window size to twice its original size. (c, f) Using the chirp Z-transform operator. The parameters are NA $= 1.4$, immersion medium: water of refractive index 1.33, polarization: circular; emission wavelength $\lambda_{\text{em}} = $ \SI{580}{\nano\meter}, voxel size \SI{80}{\nano\meter} $\times$ \SI{80}{\nano\meter} $\times$ \SI{140}{\nano\meter} and, displayed window size: $256\times 256\times 25$ pixel.}
	\label{fig:SP vs Zeropadd vs CZT}
\end{figure}

Three possible strategies can help to avoid this wrap-around effect by:
\begin{itemize}
	\item[A.] Zero-padding the in-focus plane to extend the axial depth from whereon the standing wave patterns occur (see Fig. \ref{fig:zeropad xy} and \ref{fig:zeropad xz}),
	\item[B.] Establishing an ideal absorptive boundary condition to the outside boundary and continuing the propagation by re-projecting the filtered field onto the pupil plane \cite{PointSpreadFunctions_jl},
	\item[C.] Using a chirp Z-transform (CZT) to calculate only a part of the propagated field while sapling the pupil at the finest spacing (see Fig. \ref{fig:czt xy} and \ref{fig:czt xz}).
\end{itemize}

Option A is computationally expensive since twice the initial image window size slows the calculation down by a factor $\sim$5. However, padding with zero to twice the original size may still yield unacceptable artifacts in 3D PSF calculations. Option B has the disadvantage of sacrificing a good PSF for a portion of pixels near the edge of the lateral grid of the calculation. In addition, two FFTs are required for this approach instead of only one for each propagation step. With option C, wrap-around artefacts can partially be avoided at the expense of more than twice increase of computation time. For our work, we selected option C. This can be observed in Fig. \ref{fig:czt xy} and \ref{fig:czt xz} where the standing waves are seen to be considerably reduced. The use of CZT for Fourier optics and PSF modelling is not new in the literature and has been shown to be more efficient than FFTs with zero-padding without loss of accuracy \cite{leutenegger2006fast, bakx2002efficient, smith2016simultaneous}. 

\section{Fourier-based methods for PSF calculation}\label{methods}
We present four Fourier-based methods for computing the PSF of a conventional fluorescence microscope in this section. The four methods are based on the finding of McCutchen \cite{mccutchen1964generalized} and differ how the field is propagated in free space. 

\subsection{The electric field on the $k-$sphere}
\begin{figure}[htbp]
	\centering
	\centering
	\includegraphics[width=0.75\textwidth]{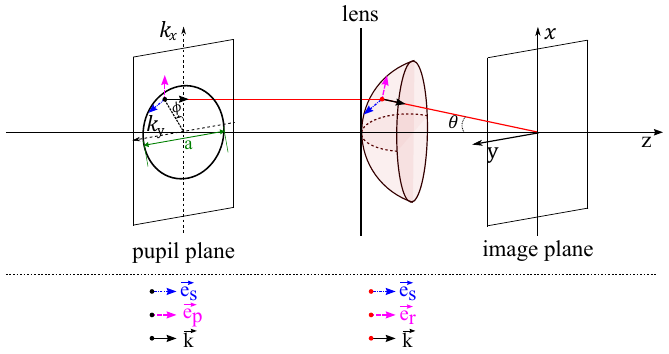}
	\caption[Vector field propagation from a circular aperture (at the pupil plane) of diameter $a$ through a high $\NA$ system.]{Vector field propagation from a circular aperture of diameter $a$ through a high NA system. $\vec{e}_s, \vec{e}_p$ are unit vectors of $s$ and $p-$polarization state, $\vec{e}_r$ represents the unit vector of the $p-$polarization after the projection on the Gaussian reference sphere.} 
	\label{fig:vector field diagram}
\end{figure}

To calculate the electric field on the $k-$sphere, we generalize the theory described in Section \ref{McCutchen pupil} from scalar to vector formulation, accounting for the polarization state of the input field. We associate each plane wave arriving at the focus with the refractive effect (``bending'') applied to the corresponding ``ray'' at the equivalent refractive locus on the Gaussian reference sphere. We assume a perfect anti-reflection coated objective lens (system) transmitting all the energy of rays. We assume a system satisfying the Abbe sine condition, which requires the beams to change direction at the Gaussian reference sphere. Knowing that the electrical field of the plane wave is a transversal wave and neglecting any axial components of the field, we thus need to project the electric field at the 2D pupil plane ($E_x, E_y$) onto a plane perpendicular to the new propagation direction, to obtain the electric field $(E_x, E_y, E_z)$ of each position on the 3D McCutchen pupil. The diagram in Fig. \ref{fig:vector field diagram} illustrates this effect.

Let $\vec{E}_i = (E_x,E_y,0)$ denote the incident polarized wave from infinity at the left side of Fig. \ref{fig:vector field diagram}. It is useful to consider a locally varying coordinate system along azimuthal ($\vec{e}_s$) and radial ($\vec{e}_p$) directions respectively. Let $\vec{E}_t$ denote the field amplitude transmitted along the wave vector $\vec{k}$ towards a point ($x,y,z$) near the focus where the field is evaluated (see Fig. \ref{fig:vector field diagram}). The unit vector corresponding to the radial component $\vec{e}_p$ is refracted by $\theta$ and becomes $\vec{e}_r$ while the azimuthal component oriented along $\vec{e}_s$ remains unchanged.

\begin{equation}
	\vec{e}_s=\begin{pmatrix}
		-\sin \phi \\
		\cos \phi\\
		0
	\end{pmatrix} , \quad \vec{e}_p=\begin{pmatrix}
		\cos \phi \\
		\sin \phi\\
		0
	\end{pmatrix}
	\Rightarrow
	\vec{e}_r=\begin{pmatrix}
		\cos \phi \cos\theta\\
		\sin \phi\cos\theta\\
		\sin\theta
	\end{pmatrix}
\end{equation}

The field amplitude distribution on the $k-$sphere oriented towards a point $(x,y,z)$ in the image plane is therefore given by:
\begin{equation}
	\vec{E}_t(x,y,z) = (\vec{E}_i \cdot \vec{e}_p)\vec{e}_r + (\vec{E}_i \cdot \vec{e}_s)\vec{e}_s.
	\label{Eq: Et McCuthen pupil}
\end{equation}

The incident wave field $\vec{E}_i$ is with a given polarization state. In the presence of refractive index mismatch in the system, a phase aberration, $\Phi$, is introduced into the system and the $p$ and $s-$polarized light are transmitted at a rate determined by the transmission coefficient of each polarization state. The amplitude field  on the McCutchen pupil is generalized as follows: 

\begin{equation}
	\vec{E}_t=[T_p(\vec{E}_i\cdot\vec{e}_P)\vec{e}_r + T_s(\vec{E}_i\cdot\vec{e}_s)\vec{e}_s] \text{e}^{ik_0\Phi},
	\label{Eq: Et McCuthen pupil with polarization}
\end{equation} 

$k_0 = 2\pi/\lem$ being the wavenumber and $T_p$ and $T_s$ are the transmission coefficients for $p$ and $s$ polarization respectively. With the assumption of an anti-reflection coated objective lens, we can further assume that the reflection coefficients are smaller than one. Omitting the phase terms in the complex transmission coefficients and add it as part of the phase term in $\Phi$, the expressions of $T_p$ and $T_s$ for $N$ number of layers in a stratified medium and $N-1$ of interfaces are therefore formulated as follows:

\begin{equation}
	T^{(N-1)}_m \approx \left(\displaystyle\prod_{j=1}^{N-1}t_m^{(j)}\right)  \text{ with } t_m^{(j)}=\frac{2\alpha p_m^j}{p_m^j+p_m^{j+1}}, \alpha = \begin{cases}
		1 & \text{if } m=s\\
		n_j/n_{j+1} & \text{if } m=p
	\end{cases},
	\label{Eq:Transmission(N-1)Approx}
\end{equation}

with $p_s^j = n_j\cos(\theta_j)$ and $p_p^j = \frac{1}{n_j}\cos(\theta_j)$, $j\in [1,N]$ and $(j)\in [1,N-1]$ \cite{born3rdEdPrinciples}.

Any additional phase or amplitude modulation can easily be introduced in the formulation of the electric field in Eq. \eqref{Eq: Et McCuthen pupil with polarization}. In the next sections, we present the different methods from this work to propagate the fields in free space.

\subsection{Slice propagation method with FFT (SP-FFT)}
This method uses the well known angular spectrum method to propagate the field in free space slice by slice  without zero-padding of the input field. Based on the Fourier slice theorem, a $z-$slice in real space corresponds to an integral of the amplitude over the axial spatial frequency $k_z$ in Fourier space. A defocus by a distance $z$ in real space corresponds to a phase shift equivalent to $k_zz$ in Fourier space. Our method here accounts for the jinc-FT trick described in Section \ref{jinc-trick} to avoid the pitfalls of the digital Fourier transform and the aplanatic correction for energy conservation. The steps for its computation is summarized in Algorithm \ref{alg:sp fft}:

\begin{algorithm}[htbp]
	\caption{Slice propagation: SP-FFT}\label{alg:sp fft}
	\begin{algorithmic}[1]
		\Require $\text{NA}_{\text{eff}}$, $\lem$, $z$, [$p_x,p_y,p_z$], [$N_x,N_y,N_z$] 
		\Ensure $[h: $ PSF intensity; $\hamp: $ complex amplitude field]  
		\State Calculate the phase aberration $\Phi$ and the transmission coefficients $T_p$ and $T_s$ for $p$ and $s$ polarization state in presence of a stratified medium with different refractive index mismatch
		\State Define the 2D incident polarized transverse wave $\vec{E}_i = (E_x,E_y)$ at the pupil plane
		\State Construct projected field components of the McCutchen pupil of uniform magnitude $\vec{E}_t(k_x,k_y)$ with the jinc-FT trick as an aperture delimiter including possible phase modification $\Phi$
        \State Apply the vectorial projection using the local coordinate systems $\vec{e}_s$ and $\vec{e}_p$ defined in Eq. \eqref{Eq: Et McCuthen pupil with polarization}   
		\State Apply the aplanatic factor for energy conversation ($\sqrt{\cos\theta}$) including the effect of projecting the McCutchen pupil ($1/\cos \theta$) leading to $\vec{E}'_t(k_x,k_y) = (1/\sqrt{\cos\theta})\cdot \vec{E}_t(k_x,k_y)$ 
		\State Propagate this pupil to real space for each desired $z-$position:
		 $\hamp(x,y,z) = 
		 \text{FFT}^{-1}_{\text{2D|xy}}\left(\vec{E}'_t(k_x,k_y)\cdot e^{ik_zz} \right)$ 
		\State Calculate the PSF intensity: $h = |\hamp|^2$
	\end{algorithmic}
\end{algorithm}

In the presence of stratified medium of different refractive indices, the effective numerical aperture is equal to $\text{NA}_\text{eff} = \min\{\text{NA}, n_j, \forall j \in [1,N]\}, N$ being the number of layers, $n_j$ the refractive index of the $j^\text{th}$ layer and NA is the numerical aperture of the objective lens as given by the manufacturer. The pixel pitch in $x,y,z$ is stored in the variables $[p_x,p_y,p_z]$ and are in the same units as the emission wavelength $\lem$. We denote $[N_x,N_y,N_z]$ the size of the grid in $x,y,z$ respectively in pixels. 

\subsection{Slice propagation method with CZT (SP-CZT)}
The SP-FFT still has its limitations due to Fourier wrap-around at higher depth as described in Section \ref{Fourier wrap-around}. The CZT is an alternative operator to reduce or totally avoid the wrap-around effect since it allows to specify the (zoomed) region after the transform implicitly applying zero-padding to the field to transform. For a $1$D signal $X_q, q\in[0,Q-1]\cap\mathbb{N}$ with $Q$ being the number of points of the signal and $\mathbb{N}$ the set of natural numbers, the Z transform $\tilde{X}_z, z \in \mathbb{C}$ is given as follows:
\begin{equation}
	\tilde{X}_{z_m} = \text{CZT}(X_q) = \sum_{q=0}^{Q-1}X_qZ_m^{-q},
	\label{Eq:original czt eq}
\end{equation}

where $Z_m=AW^{-m}, m\in\mathbb{N}$ is a spiral path ($Z-$path) in the complex plane with $A$ being the starting point and $W=\exp(-i\Delta\beta)$ the ratio of two consecutive points with a given angular increment phase $\Delta\beta$. For $A=1$ and $\Delta\beta=2\pi m/Q, Z_m$ is computed over an unit circle and the CZT operation becomes a discrete FFT. To zoom the signal $X_q$ in by a scalar factor $c$, $A=\exp(-i\pi/c)$ and $W=\exp(-i2\pi/(Qc))$ \cite{rabiner1969chirp}. Therefore, Eq. \ref{Eq:original czt eq} can be expressed in terms of convolution as follows \cite{leutenegger2006fast}:

\begin{equation}
	\tilde{X}_{z_m} = W^{m^2/2}\text{FFT}^{-1}\left(\text{FFT}\left(X_qA^{-q}W^{q^2/2}\right)\cdot\text{FFT}\left(W^{q^2/2}\right)\right).
    \label{Eq:czt formula}
\end{equation}

The inverse CZT of a signal $\tilde{X}_{z_m}$ in a frequency-domain representation is defined as the complex conjugate of the CZT of the complex conjugate $\tilde{X}_{z_m}^*$ of $\tilde{X}_{z_m}$ within some scaling factor for a CZT operating on a unit circle \cite{frickey1995using}. 

\begin{figure}[htbp]
	\centering 
	\includegraphics[width=0.5\textwidth]{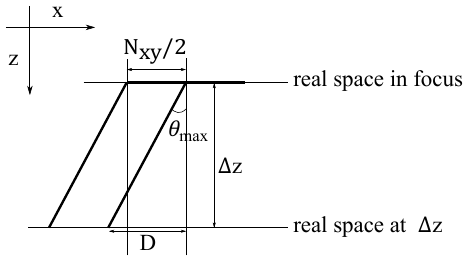}
	\caption{Diagram of the wave propagation in real space for calculating the zoom factor $c$.}
	\label{fig:czt zoom factor}
\end{figure}

In this approach, we calculate a zoom factor $c$ to perfectly fit the pupil near the edge of the array to transform, corresponding to Nyquist sampling of the amplitude in real space. By doing so, we use all the available lateral space in Frequency space. We calculate the field in Frequency space according to Step 1 to 5 in Algorithm \ref{alg:sp fft}. If necessary, we can calculate the field on the pupil plane on a finer grid to obtain an even finer sampling in Fourier space. We use the inverse CZT to zoom-in (\textit{i.e.} zoom out in the pupil plane) using the zoom factor $c$ to obtain the corresponding real space field with the desired sampling. To minimize wrap-around, the appropriate zoom factor $c$ is calculated such that the lateral window size of the calculated PSF is slightly bigger or equal to the lateral dimension of the PSF at the highest depth position $\Delta z$ from the focus. The maximum angular aperture of the optical emission determines the maximum angle at which an oblique emission will propagate. We denote D the lateral radius of the propagated beam at $\Delta z$ (see Fig. \ref{fig:czt zoom factor}).

The factor $c$ is therefore calculated as $c=(D+N_{xy}/2)/(N_{xy}/2)$, where $\tan\theta_{\text{max}}=D/\Delta z$ in real space and $\tan\theta_{\text{max}}=k_{xy}/k_z$ in Fourier space with $\theta_{\text{max}}$ being the maximal angular aperture and $N_{xy} = [N_x, N_y]$ the number of pixels in the in-focus $xy$-plane (see Fig. \ref{fig:czt zoom factor}). If the number of pixels in $x$ and $y$ directions are not equal, the zoom-in factor $c$ will also be different in both directions or can be set to be the minimal value. Given that the in-focus plane is calculated over $N_{xy}$ pixels, the PSF may require $N'_{xy} = 2(N_{xy}/2+D)$ lateral grid. 

We modify Algorithm \ref{alg:sp fft} to introduce the wave propagation using CZT method and the result is summarized in Algorithm \ref{alg:sp czt}.

\begin{algorithm}[htbp]
	\caption{Slice propagation: SP-CZT}\label{alg:sp czt}
	\begin{algorithmic}[1]
		\Require $\text{NA}_{\text{eff}}$, $\lem$, $z$, [$p_x,p_y,p_z$], [$N_x,N_y,N_z$] 
		\Ensure $[h: $ PSF intensity; $\hamp: $ complex amplitude field]  
		\State Calculate the required window size $N'_\text{xy}$ and the zoom-in factor $c$ 
		\State Compute Step 1 to 5 in Algorithm \ref{alg:sp fft} under the required window size $N'_\text{xy}$ and lateral pixel pitch $p'_\text{xy} = c\times [p_x, p_y]$ to obtain the zoomed field $\vec{E}^{'}_t$ on the McCutchen pupil
		\State Calculate the propagator $e^{ik_zz}$ 
		\State Propagate the field in real space using the CZT$^{-1}$ and by applying the zoom-out factor $c$ calculated in Step 1 to obtain $\hamp(x,y,z) = \text{CZT}^{-1}_{\text{2D|xy}}\left(\vec{E}^{'}_t \cdot e^{ik_zz} \right)$ 
		\State Calculate the PSF intensity: $h = |\hamp|^2$
		\State Crop $h$ and $\hamp$ to the desired initial window size [$N_x,N_y,N_z$] 
	\end{algorithmic}
\end{algorithm}

\subsection{Fourier-shell interpolation method (F-Shell)}
This method aims at representing the useful part of the McCutchen pupil directly in $3$D-Fourier-space and projecting the two-dimensional pupil modifications (the aplanatic factor and vectorial projection effects) onto this three-dimensional shell. The difficulty is that the shell, at each integer $[k_x,k_y]$ position, has a non-integer $k_z$ position leading to an interpolated representation of $k_z$ in Fourier space. 

\begin{figure*}[htbp]
     \centering
     \begin{subfigure}[h]{0.49\textwidth}
     	\centering
     	\includegraphics[width=\textwidth]{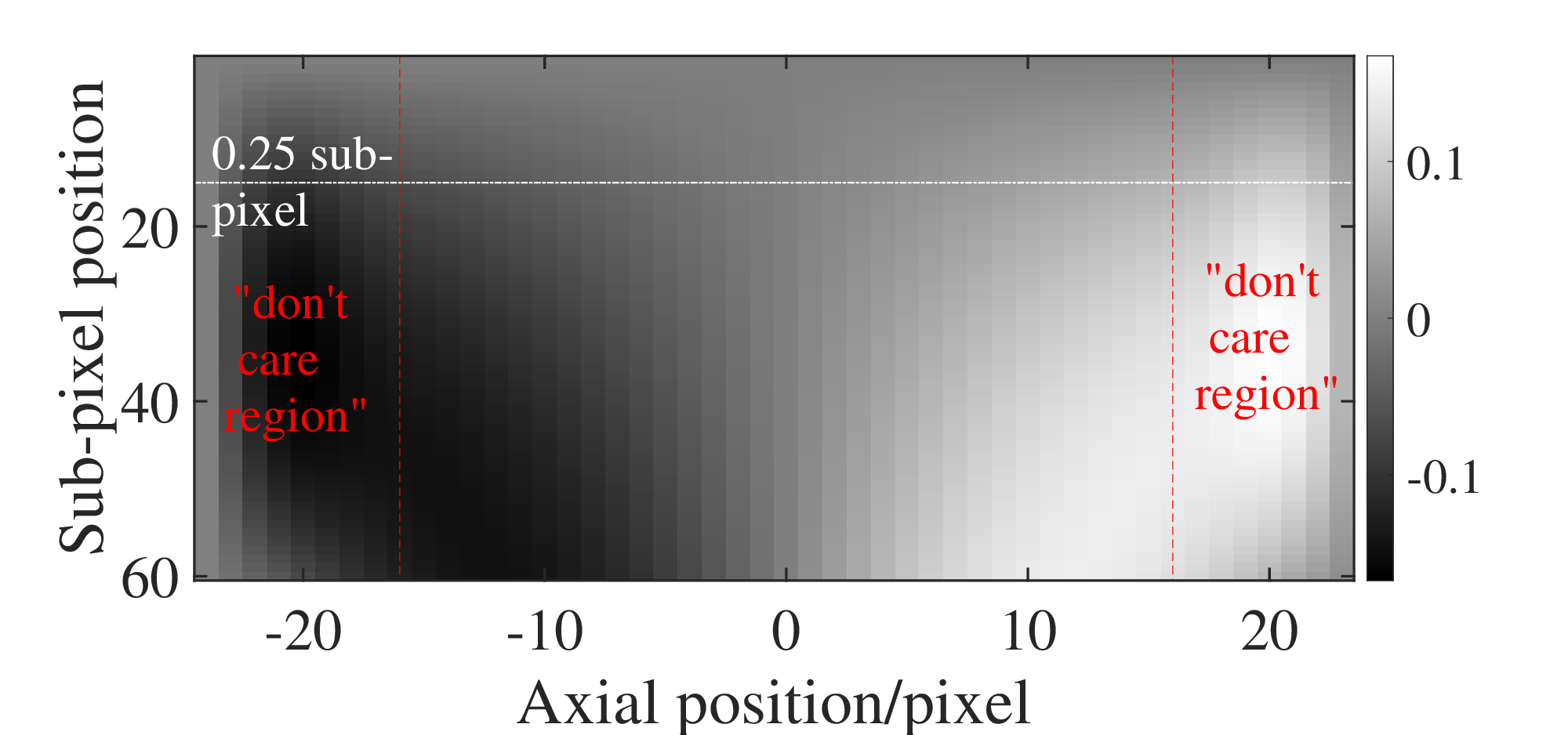}
     	\caption{\label{fig:phase shift table}}
     \end{subfigure}
     \begin{subfigure}[h]{0.49\textwidth}
         \centering
         \includegraphics[width=\textwidth]{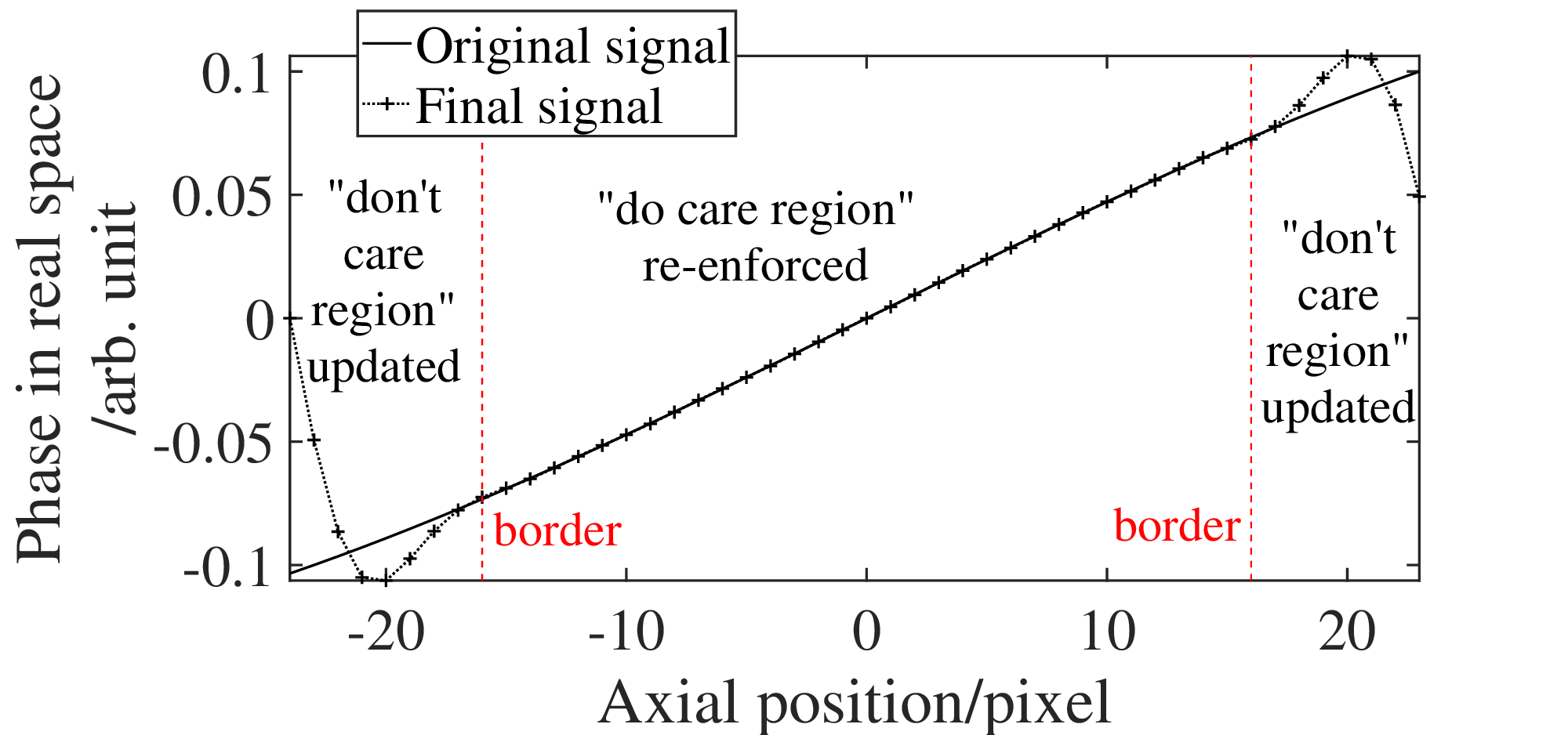}
         \caption{\label{fig:coeff matrix rspace}}
     \end{subfigure}\\
	 \begin{subfigure}[h]{0.49\textwidth}
	 	\centering
	 	\includegraphics[width=\textwidth]{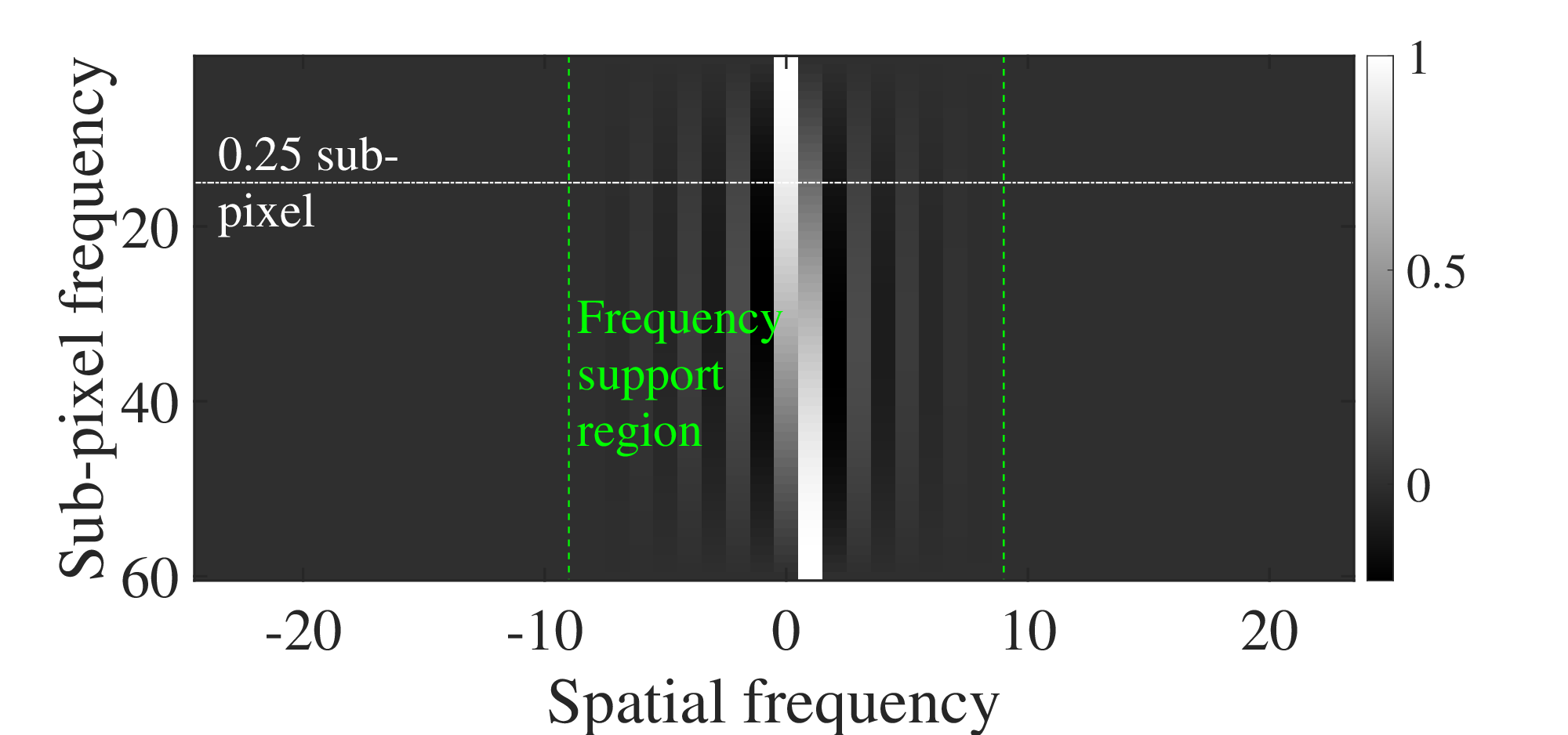}
	 	\caption{\label{fig:interpolator table}}
	 \end{subfigure}
     \begin{subfigure}[h]{0.49\textwidth}
        \centering
         \includegraphics[width=\textwidth]{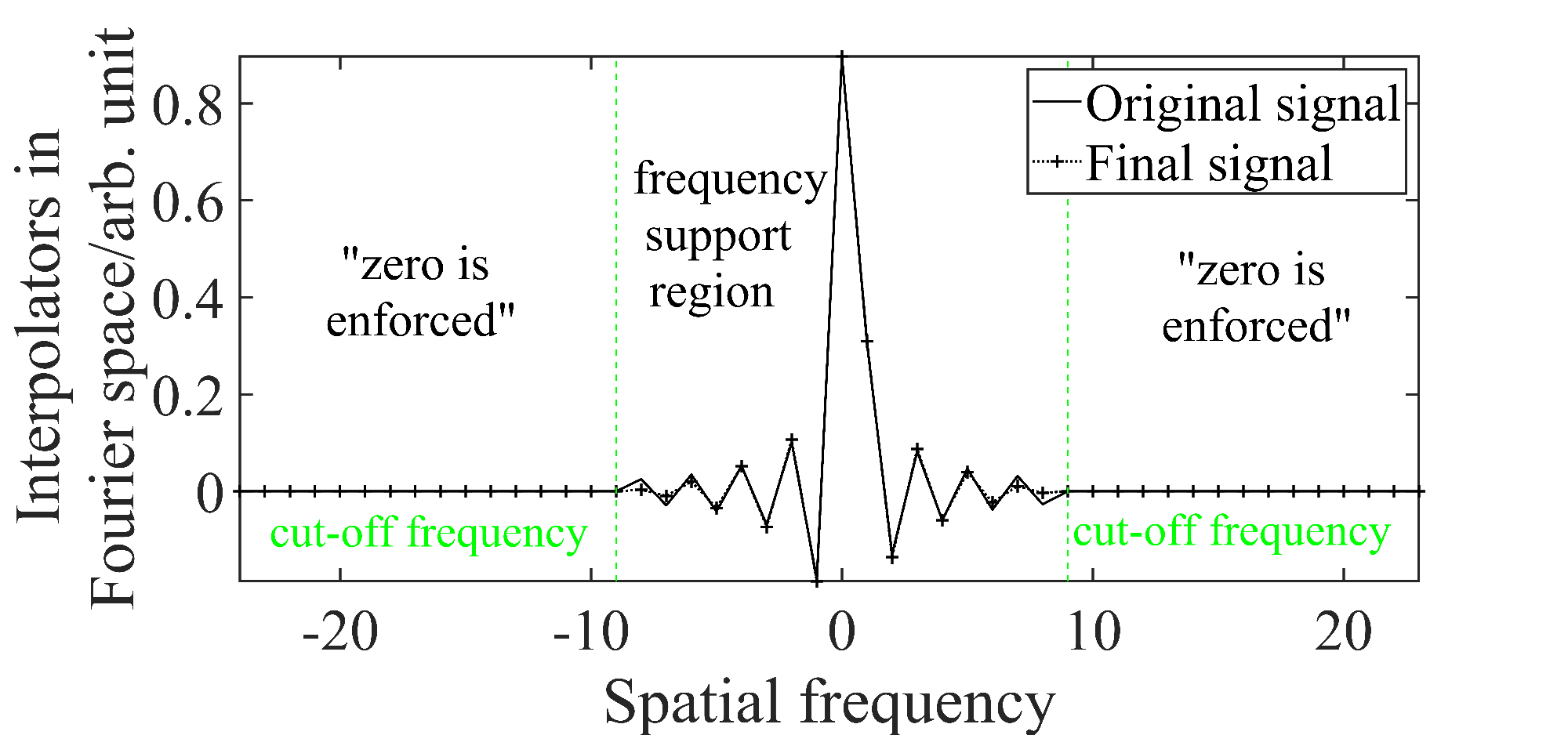}
         \caption{\label{fig:coeff matrix fspace}}
     \end{subfigure}
     \caption{(a) Phase shift in the ideal wave $\exp(2\pi i k_zz)$. (b) Phase at a sub-pixel $0.25$, indicated by the horizontal white line in (a). (c) Interpolation table in Fourier space containing the interpolator coefficients at $60$ different sub-pixel positions. (d) Interpolation coefficients in Fourier space along the $0.25$ sub-pixel indicated by the horizontal white line in (a) and (c).}
     \label{fig:VS coeff matrix}
\end{figure*}

As a credible representation of such a non-integer $k_z$ would require essentially the full $k_z$-range. To keep the computation efficient, an appropriate compromise was therefore made. We calculate a table of interpolation kernels ("interpolators"), each of them applicable for a specific (small range of) sub-pixel shifts. We aim to represent only the central part of the corresponding real-space representation as faithfully as possible and label a border region as ``don\textquotesingle t care'' region (see Fig. \ref{fig:coeff matrix rspace}). This ``don\textquotesingle t care'' region is limited by a chosen factor $b_{\text{reg}}$ (here it is chosen to include 8 pixels from both edges). The part of real space within this border region is iteratively updated, while the central part is forced to the expected values in each iteration in this iterative Fourier transformation algorithm (IFTA) to obtain the interpolation kernels. 

In addition, a pre-defined cut-off frequency $n_{k_{\text{cut-off}}}$ is chosen. This cut-off frequency limits the number of interpolation coefficients given by $n_z = 2n_{k_{\text{cut-off}}}+1$, which can be used to fill the voxels along $k_z$ in Fourier-space adjacent to the one nearest to the non-integer $k_z (k_x,k_y)$ position of the McCutchen pupil. The required interpolation coefficients are generated with the help of the IFTA \cite{alsaka2018iftacomparison}. We denote $n_{\text{subpix}}$ the number of sub-pixel positions along $k_z$. As initialization, ideal $\exp(2\pi i k_zz)$ waves are generated in real space corresponding to the respective sub-pixel frequencies in Fourier space. The ideal waves are then Fourier-transformed and only $n_z$ interpolator values are kept while all others are set to zero. The result is transformed back to real space, where the central area is replaced by the original perfect waves, but the ``don\textquotesingle t care'' region is not touched. This is repeated over $N_\text{iter}$ iterations (typically $500$ times) until convergence. The so-generated interpolation table of size $n_z\times n_\text{subpix}$ is stored for later use (see Fig. \ref{fig:interpolator table}). 

In the example presented here, $n_{k_{\text{cut-off}}} = 8$ leading to $n_z = 17$ interpolation coefficients to be determined. An interpolation table of $n_{\text{subpix}} = 60$ sub-pixel positions of the 17 complex valued coefficients is pre-computed via IFTA. The inner part in the real space regime which represents the ``do care region''  is about \SI{66}{\percent} of the given $z$-range. A typical example for the offset of \SI{0.25}{pixels} is shown in real and Fourier space in Fig. \ref{fig:coeff matrix rspace} and \ref{fig:coeff matrix fspace} respectively, overlayed with the ideal subpixel wave (solid line which corresponds to the legend `Original signal'). The border of the ``don\textquotesingle t care region'' is indicated by the dashed vertical lines. A real space representation of the full interpolation table is shown in Fig. \ref{fig:phase shift table} with the ``don't care region'' also indicated by the vertical red dashed lines.

The size of the border factor $b_\text{reg}$ (in pixels) and the cut-off frequency $n_{k_\text{cut-off}}$ defining the number of interpolation coefficients $n_z$ should be roughly the same. If the ``don't care region'' is far bigger than the ``do care region'', fewer non-zero interpolation coefficients are needed in Fourier space, speeding up the algorithm in the application phase. However, the integrated intensity of the PSF is less uniform, increasing the violation of the missing cone  property, which will be further discussed in Section \ref{comparison}. A small ``don't care region'' on the other hand can lead to inaccuracies inside the ``do care region'' if the number of interpolation coefficients is kept small. The principle for the PSF generation using this method is summarized in Algorithm \ref{alg:vs}.

\begin{algorithm}[htbp]
	\caption{Fourier shell interpolation (F-Shell)}\label{alg:vs}
	\begin{algorithmic}[1]
		\Require $\text{NA}_{\text{eff}}, \lem, z, [p_x,p_y,p_z], [N_x,N_y,N_z], b_\text{reg}, n_{k_\text{cut-off}}, n_\text{subpix}, N_\text{iter}$ 
		\Ensure $h$: PSF intensity, $\hamp$: complex amplitude field  
		\State Generate the McCutchen pupil projections $\vec{E}'_t(k_x,k_y)$ as described in Algorithm \ref{alg:sp fft}, Step 1 to 5, using the jinc-FT trick as an aperture delimiter 
		\State Calculate $k_z (k_x,k_y)$ for every pixels within the pupil and round it to the nearest $1/n_{\text{subpix}}$ subpixel $k_z$  position
		\State Generate a 3D index of the full $k-$sphere 
		\State Store the interpolation coefficients for future use
		\State For each PSF to calculate: Write the field $\vec{E}'_t$ calculated in Step 1 with the appropriate interpolation kernel for the sub-pixel at the 3D index position of the $k-$sphere
		\State Perform a 3D FFT of the result from Step 5 to obtain the three-dimensional field distributions $\hamp$ (with expected errors in the ``don\textquotesingle t care region'')
		\State Calculate the PSF intensity: $h = |\hamp|^2$
	\end{algorithmic}
\end{algorithm}

This PSF calculation method (step 5) can be performed fast and memory efficient as a single access operation in MATLAB by exploiting its indexed addressing capabilities. In this way, the complex-valued $2$D pupil can be rapidly written into the appropriate 3D Fourier space region with the optimized interpolation coefficients as described above and the ``don\textquotesingle t care'' region can be later removed by cropping. The required $k_z$-range can be kept to a minimum. 

\subsection{Sinc-R method (Sinc-R)}

This method is derived using the knowledge that the  three-dimensional Fourier transform of a complete spherical shell is a $\sinc (k_0|r|)$ function, $r$ the spatial radial coordinate. To compute the $k-$sphere representation, in analogy to the jinc-trick, we apply the three-dimensional FFT to a $\sinc (k_0|r|)$ calculated in real space. The sinc-shell method is described in Algorithm \ref{alg:sincr}.

\begin{algorithm}[htbp]
	\caption{Sinc-R method}\label{alg:sincr}
	\begin{algorithmic}[1]
		\Require $\NA_{\text{eff}}, \lem, r, [p_x,p_y,p_z], [N_x,N_y,N_z] $
		\Ensure $[h: $ PSF intensity; $\hamp: $ complex amplitude field]  
        \State Set the calculation grid to be 1,25 times bigger in both direction $x$ and $y$ than the desired grid 
		\State Generate a $\sinc (k_0|r|), r$ being the radial coordinate, amplitude distribution in three dimensions in real space over the region described in Step 1
		\State Perform an appropriate DampEdge to the edge region of the image (e.g. \SI{5}{\percent} on each side of the image border) 
		\State Compute the 3D FFT of the distribution in Step $3$
		\State Keep only the (positive frequency) $k_z$-range, which contains valid (non-zero amplitude ) $\vec{k}$ vectors. This cropping in Fourier space changes the $z$-sampling and causes a phase ramp in real space. Note that the latter does not affect the intensity values. For a given sampling of the final intensity PSF, these steps can be adjusted accordingly.
		\State Calculate the three components of the field in the McCutchen pupil following Step 1 to 4 of Algorithm \ref{alg:sp fft}
		\State Pre-compensate for the projection factor $1/\cos\theta$ of the sinc-shell by multiplying the field with $\cos\theta, \theta$ being the angular aperture that is previously defined. Apply the factor $1/\sqrt{\cos\theta}$ as described in Step 5 of Algorithm \ref{alg:sp fft} 
		\State Multiply the resulting spectrum from Step 7 with the sinc-shell from Step 5 
		\State Perform a 3D FFT for each of the field components to obtain the sought-after field components $\hamp$ in real space
		\State Calculate the PSF intensity: $h = |\hamp|^2$
		\State Extract the field within the desired window $[N_x, N_y, N_z]$
	\end{algorithmic}
\end{algorithm}

This method has the attractive property that it does not suffer from the Fourier wrap-around effect since the field was directly generated in real-space.  A disadvantage is each step has to be performed observing Nyquist sampling along $k_z$ for the full field including its $z$-propagation. This method is also not readily applicable to a single slice (in or out-of-focus). 
\section{Quantitative PSF comparison}\label{comparison}

%
%
%

\subsection{Comparison methods}\label{Sec:comparison methods}
We compare the four models described in this project with state-of-the-art (see below) PSF models. The Richards and Wolf (RW) model was chosen as a gold standard for the comparison because it is a well accepted model of the field in an aplanatic system and its accuracy relies on the computation of the three integrals in its analytical expression. We denote the gold standard by RW-GS. To avoid interpolation problems in the calculated PSF especially near the focus position, we sample in our RW-GS $10\times$ higher than the standard sampling (see below) in $x$ and $y$. This oversampled RW-GS is then subsampled by considering only every $10^\text{th}$ pixel along $x$ and $y$ to correspond to what we choose as ``standard sampling''. The standard sampling corresponds to a voxel size of $83\times 83\times 100$ \SI{}{\nano\meter^3} and a calculation grid of $127\times 127\times 65$ pixels. We only consider $90\%$ of the calculation grid, the central $115\times 115$ pixels in each slice in our quantitative comparison to discount any artefacts that may be present at the edge of the border of the calculation grid. We assume a water objective lens of numerical aperture equal to 1.2 and an emission wavelength of \SI{510}{\nano\meter} in the simulation.

We choose four state-of-the-art commonly used PSF models to compare our methods to. The first model is the scalar PSF based on the work of Gibson and Lanni \cite{gibson1989diffraction}, and further developed by Li et al. \cite{li2017fast}. This technique calculates the PSF fast by using a combination of Bessel series. We denote this model GL in our comparison. The second and third models are a scalar and vector PSF as described in \cite{aguet2009super}, computed using a numerical integration based on Simpson's rule. We denote sPSF the scalar PSF and vPSF the vector PSF. The GL, sPSF and vPSF firstly compute a $rz$-map, $r$ and $z$ being the radial and axial coordinate, of the PSF. The radial symmetry is used to fill the whole volume linearly interpolating in the $rz-$map. The last state-of-the-art PSF to compare to is the vector PSF calculated with the Richards and Wolf $3$D optical model in the PSFGenerator toolbox in \cite{kirshner20133}. We denote this PSF by PSFGen. The four models described in this manuscript are distinguished from those state-of-the-art using an asterisk: SP-FFT*, SP-CZT*, F-Shell* and Sinc-R*.  GL, sPSF, vPSF and PSFGen are not centred at the same centre position as our PSFs models for even sizes if the calculation window. We therefore choose odd-sized calculation window grids throughout this document. SP-CZT* is however calculated on even grids and is further cropped to obtain the same window size as the other PSF models since the CZT function does not center the PSF at the same position as the other models for odd-sized grids.  

In Section \ref{comparison:intensity}, 
we investigate and quantify the accuracy of the intensity values of the PSFs. We use as error metric the relative square error (RSE) of a particular calculation with the gold standard RW-GS to quantify the performance-error of each simulation in comparison to the RW-GS 

\begin{equation}
    \text{RSE} = \frac{\sum_{x,y,z}|\text{PSF}_\text{RW-GS} - \text{PSF}_\text{sim}|^2}{\sum_{x,y,z}|\text{PSF}_\text{RW-GS}|^2}, 
\end{equation}
with $\text{PSF}_\text{sim}$ being the simulated PSF to compare with the RW-GS PSF. 

We further evaluate to what extent each PSF model satisfies the conservation of energy by quantifying the violation of the missing cone in Section \ref{comparison:missing cone}. 

As the models described in this work are Fourier-based, the effect of any remaining FFT wrap-around effects needs to be quantified to conclude on their accuracy. To achieve this for each PSF model, we calculate the error $\epsilon$ between a reference PSF, $h_\text{ref}$ computed in a window demonstrated to contain least standing waves due to Fourier wrap-around, and the PSF calculated with the same method in a different calculation grid $W$. Details on this approach are given in Section \ref{comparison:wrap-around}. The error $\epsilon$ is calculated using Eq. \ref{Eq:epsilon error}. It quantifies the amount of standing waves due to Fourier wrap-around in a given simulated PSF.
\begin{equation}
    \epsilon = \sum_{x,y,z}|h_\text{ref} - h_W|
    \label{Eq:epsilon error}
\end{equation}

Finally, we present the computation cost of each method in Section \ref{comparison:computation time}.  

\subsection{Intensity profiles of the PSFs}\label{comparison:intensity}
\begin{figure}[htbp]
	\centering
	\includegraphics[width= 1.0\textwidth]{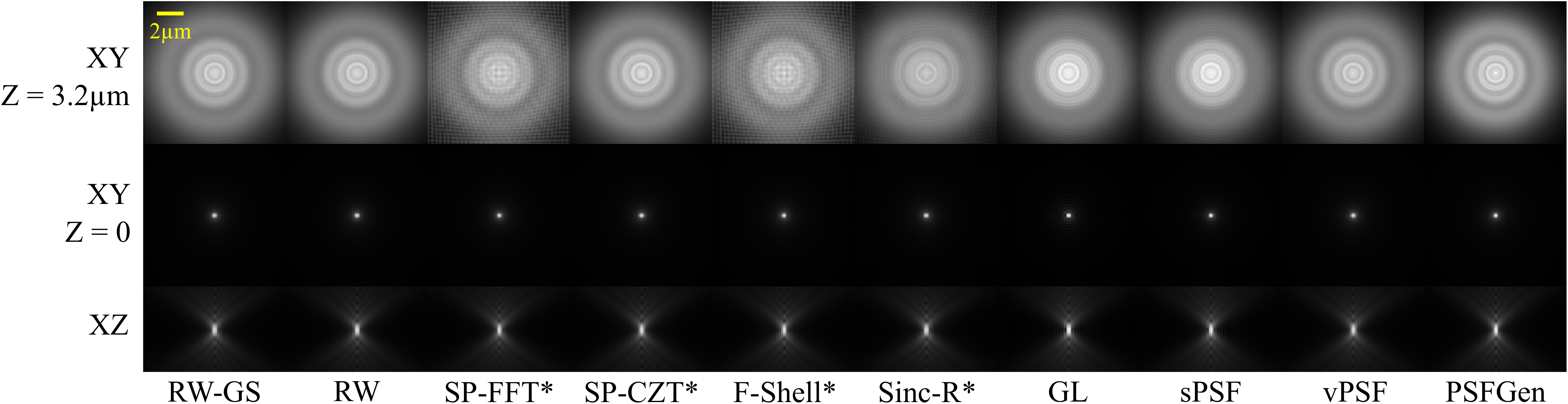}
	\caption{Display at $\gamma = 0.4$ of the $xy-$slices of the PSFs at a distance of $z = $ \SI{3.2}{\micro\meter} to the focal plane (first row) and in-focus at $z = $ \SI{0}{\micro\meter} (second row) and the central $xz-$cross section of the PSFs. RW-GS refers to the Richards and Wolf gold standard PSF, RW to the Richards and Wolf PSF under a standard sampling, Sinc-R* to the PSF derived from Algorithm \ref{alg:sincr}, PSFGen to the vector PSF in the PSFGenerator toolbox at the best accuracy \cite{kirshner20133}, SP-FFT* to the method described in Algorithm \ref{alg:sp fft}, SP-CZT* to the method employing Algorithm \ref{alg:sp czt}, F-Shell* to the Fourier shell interpolation method in Algorithm \ref{alg:vs}, GL to the scalar PSF based on the Gibson and Lanni model \cite{li2017fast}, sPSF and vPSF to the scalar and vector PSF from \cite{aguet2009super}.} 
	\label{fig:psfsall}
\end{figure}

Fig. \ref{fig:psfsall} presents $xy-$slices of the PSFs at $z = $ \SI{3.2}{\micro\meter} from the focal position and in-focus ($z = $ \SI{0}{\micro\meter}) and the $xz-$cross sections of the PSFs at $y = $ \SI{0}{\micro\meter}. Severe standing waves are observed at the $z-$depth equal to \SI{3.2}{\micro\meter} for the SP-FFT* and F-Shell* methods. The smallest region closest to the centre of the image in the Sinc-R* at \SI{3.2}{\micro\meter} also does not appear to be perfectly circular.
 
\begin{figure}[htbp]
\centering
\insertfigure{0.48}{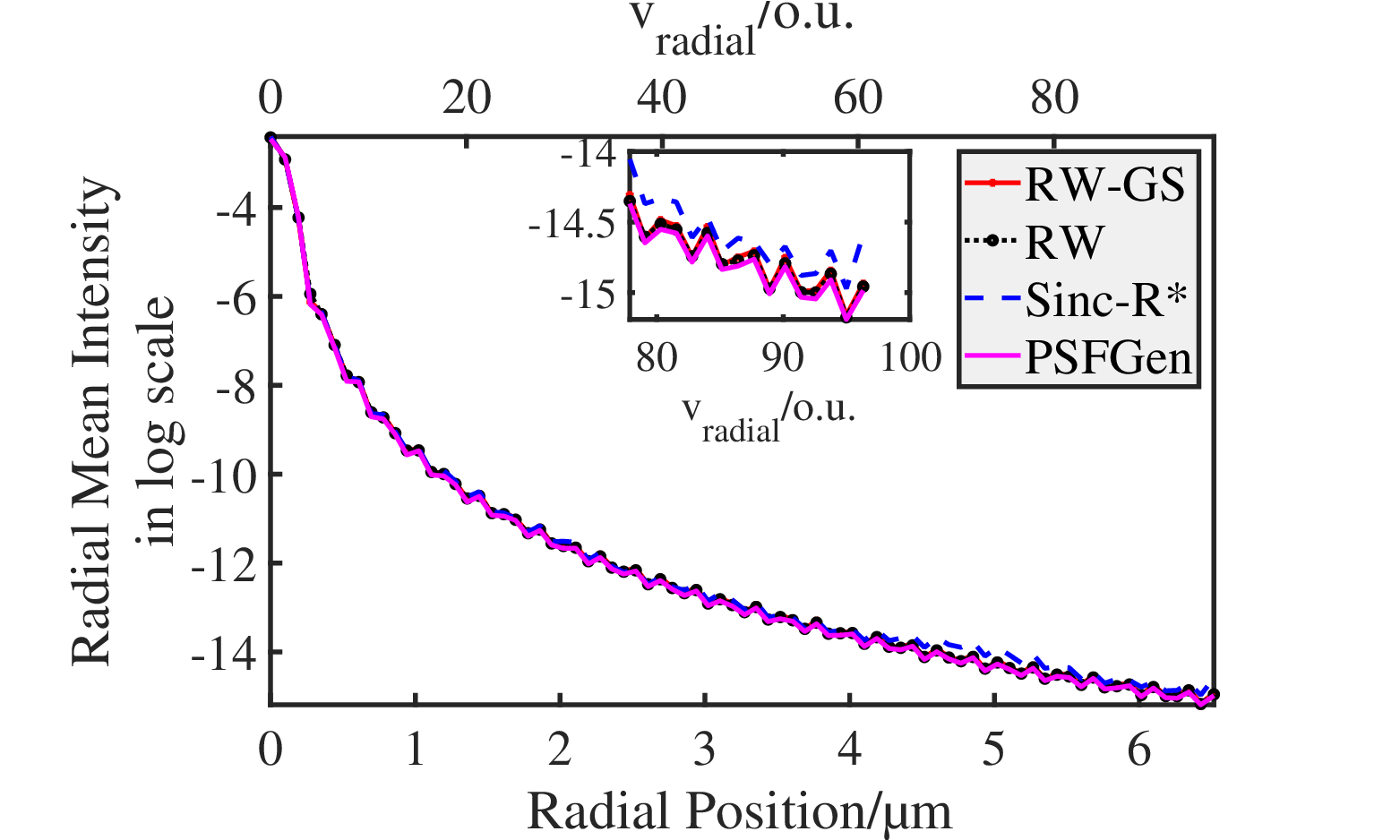}{fig:mean int gs_rw_sr_psfgen}
\insertfigure{0.48}{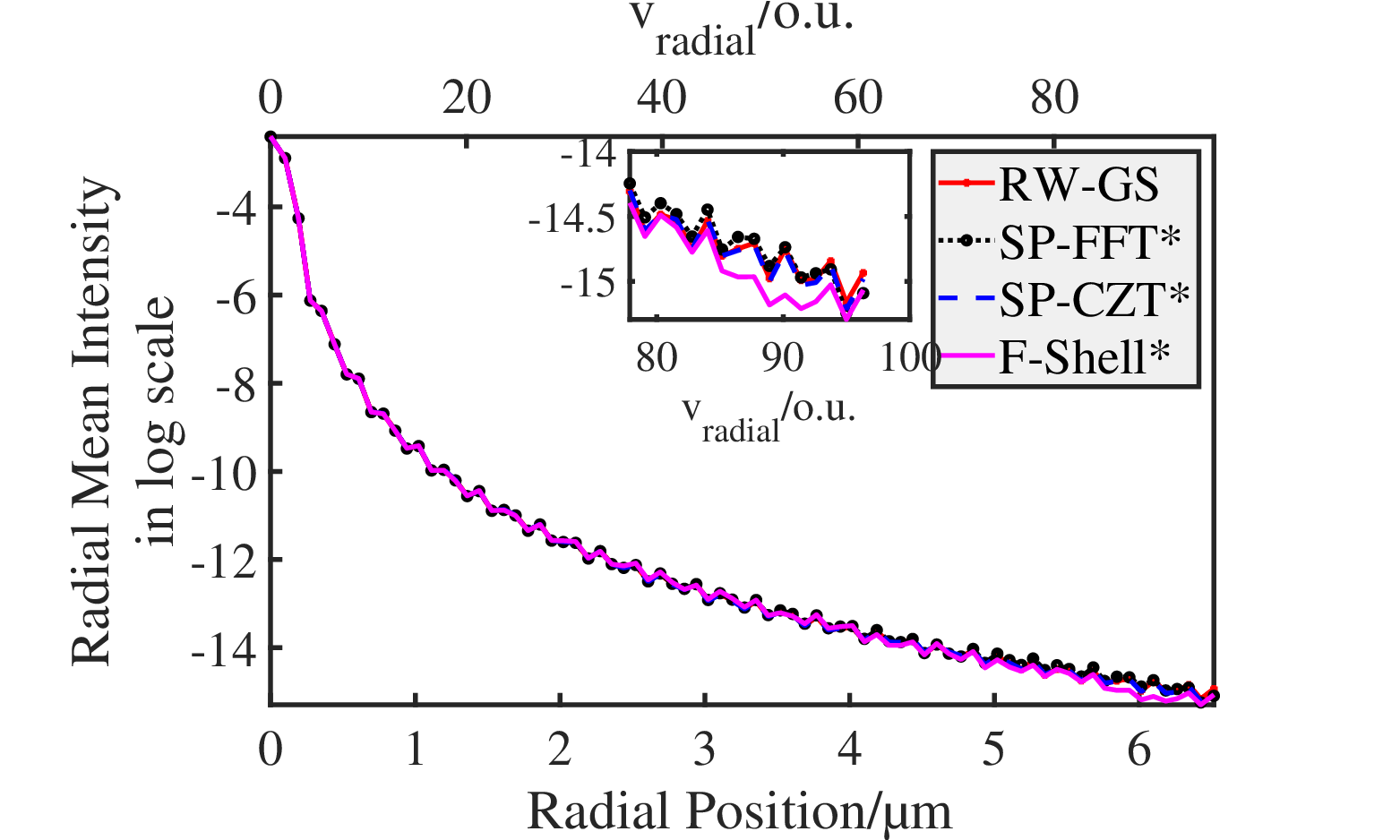}{fig:mean int gs_sp_czt_vs}
\insertfigure{0.48}{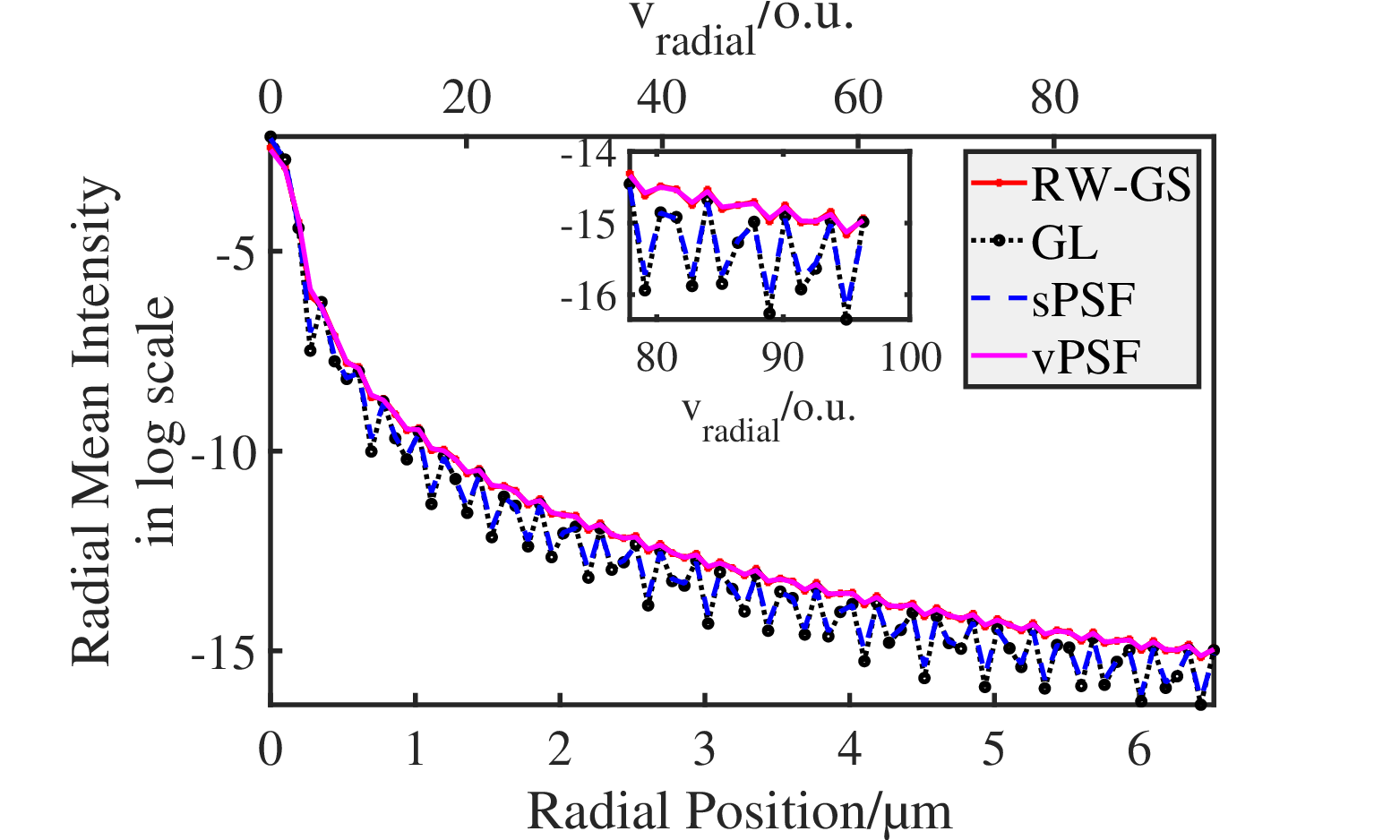}{fig:mean int gs_gl_spsf_vpsf}

	\caption{Radial mean profiles, in logarithmic scale, of the PSFs at the focal plane.}
	\label{fig:radial mean in-focus plane}
\end{figure}

To observe finer differences between our chosen gold standard RW-GS and the intensity profiles of the PSFs of various methods, we display in Fig. \ref{fig:radial mean in-focus plane} the radial mean intensity in log scale at the focus position. We observe in Fig. \ref{fig:mean int gs_gl_spsf_vpsf} a clearly visible discrepancy between the scalar PSF models (GL and sPSF) and vector models (RW-GS and vPSF). PSFs calculated by vPSF and PSFGen fit the profile of the RW-GS at high precision. Sinc-R* PSFs departs from the profile of the RW-GS beyond a distance of about half micron from the centre. The SP-FFT*, SP-CZT*, and F-Shell* methods are slightly inaccurate near the edge of the radial position.
 
We compute the relative local error map between the simulated PSFs, $\text{PSF}_\text{sim}$, relative to the RW-GS using the following equation Eq. \eqref{Eq:Err local} to visualize the error in the volume PSF and to indicate how precise the local prediction is.
\begin{equation}
\text{Err} = \frac{|\text{RW-GS} - \text{PSF}_\text{sim}|}{|\text{RW-GS}|}
\label{Eq:Err local}
\end{equation}

\begin{figure}[htbp]
	\centering
	\includegraphics[width= 1.0\textwidth]{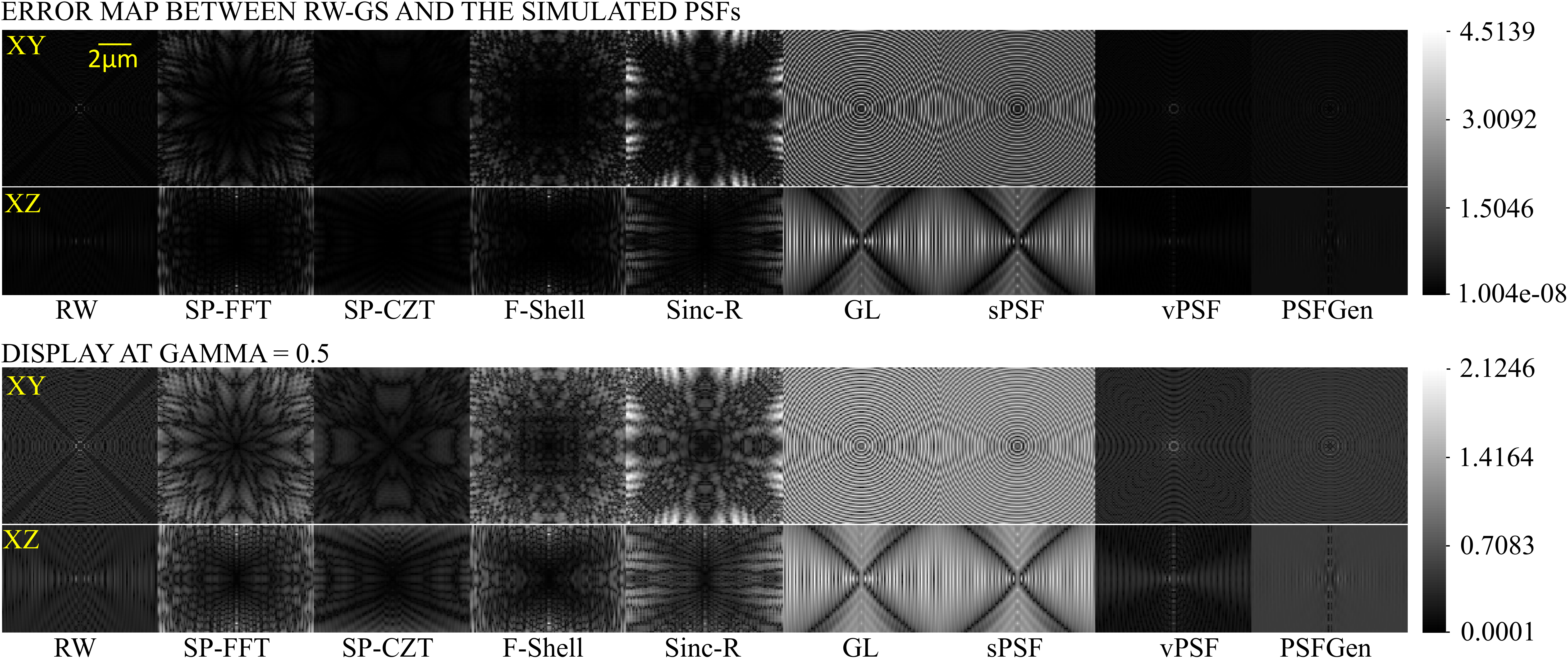} 
	\caption{Error map of the simulated PSFs relative to the highly sampled gold standard RW-GS.}
	\label{fig:error map}
\end{figure}

This map represents the absolute of the difference between the simulated PSFs and the RW-GS relative to the intensity values of the RW-GS at each pixel of the volume PSFs. The map for each of the PSFs is displayed in Fig. \ref{fig:error map}. 

It is noticed from the error map that the error in the Fourier-based models are more dominant towards the edge of the grid while the error is less significant closer to the optical axis $z$. This error in the Fourier-based models is due to the FFT and CZT operation. A discontinuity is observed in the error of the RW with normal sampling along the two diagonals of the grid. This is due to the mapping of the $rz-$map to form the whole volume PSF. The error near the axial axis is also higher than the error at different position of the grid for the RW due to its sampling. A discontinuity is observed in the $xz-$cross section of the scalar PSFs GL and sPSF where the angular aperture $\theta$ is maximal. This discontinuity is not observed in the vector vPSF, which represents a vector formulation of the sPSF. The error in vPSF is more concentrated near the region closer to the axial axis, similar to the case of the RW. 

\begin{figure}[htbp]
\centering
\insertfigure{0.48}{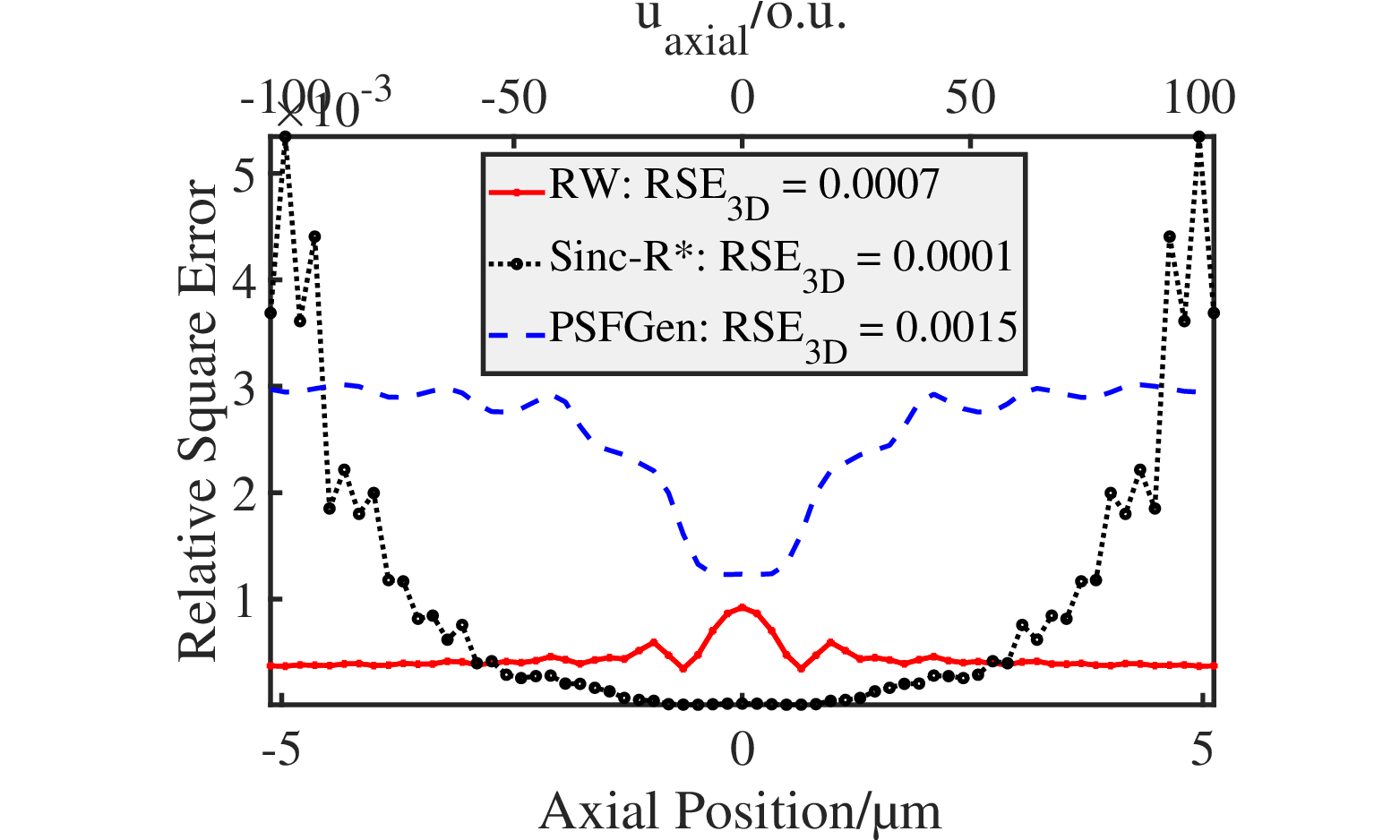}{fig:MRE RW-SR-PSFGEN}
\insertfigure{0.48}{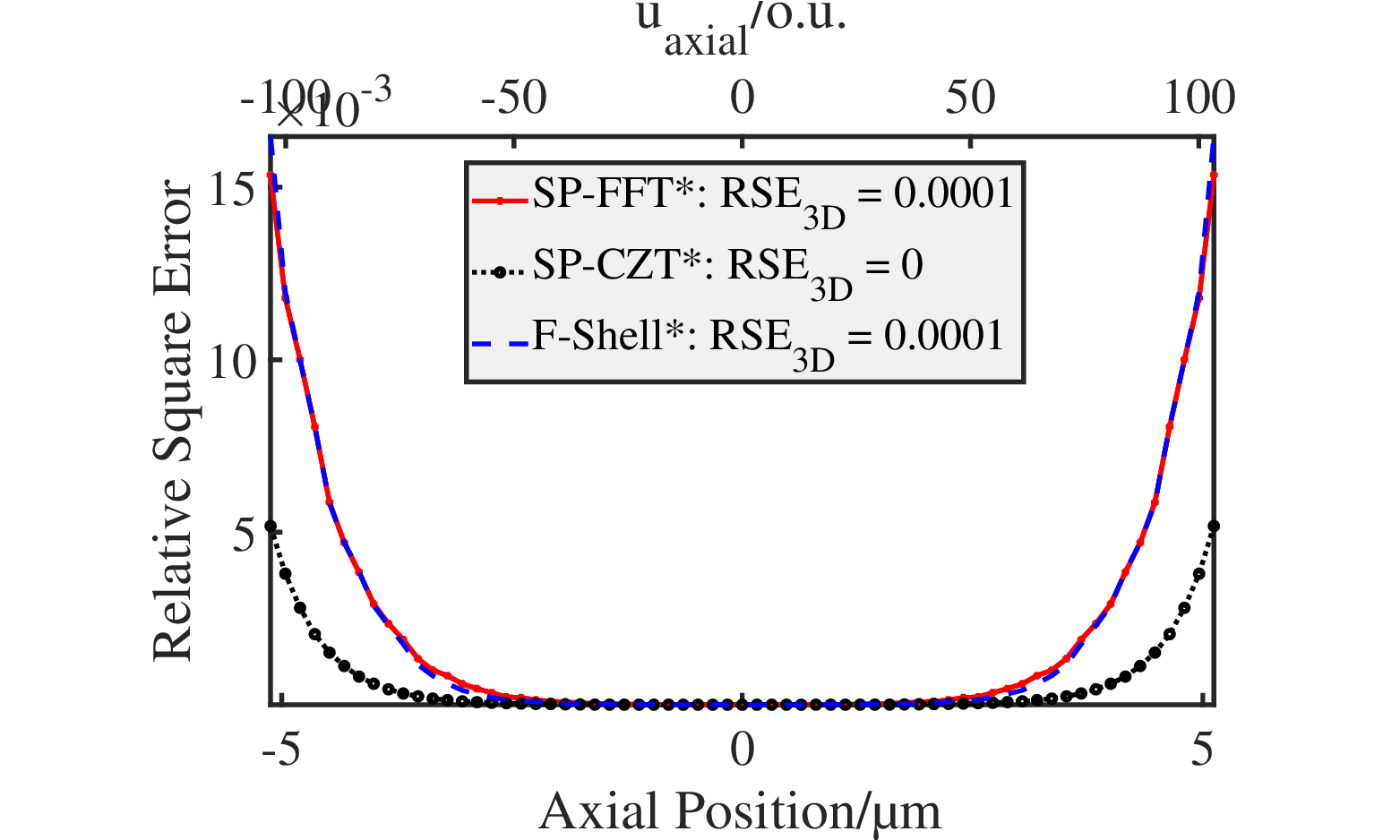}{fig:MRE RW-SP-CZT-VS}
\insertfigure{0.48}{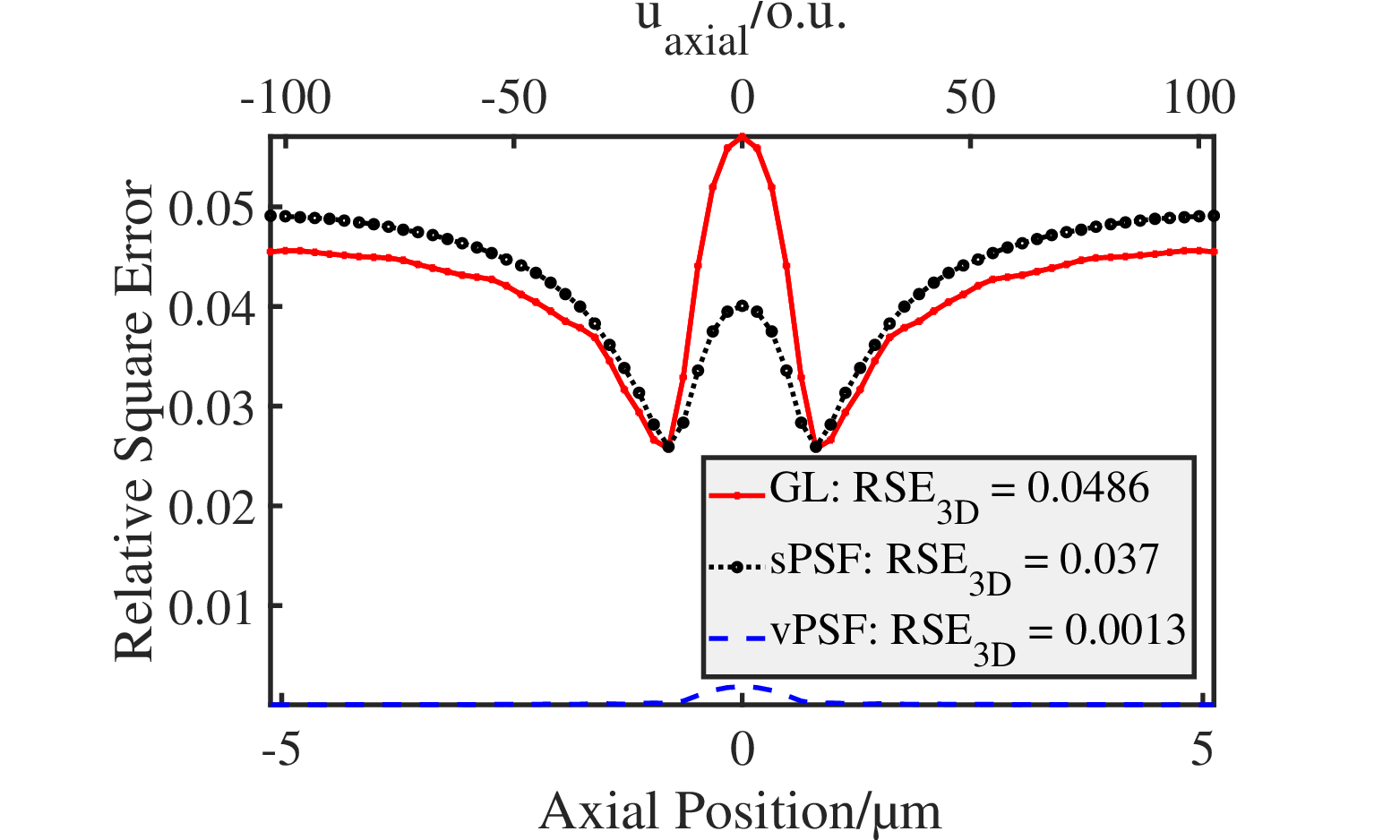}{fig:MRE RW-GL-sPSF-vPSF}

	\caption{Relative square error (RSE) between the gold standard RW-GS PSF and PSFs models.}
	\label{fig:mre gs with models}
\end{figure}

To conduct further investigation of the error in each model, we compute the RSE to the gold standard per slice ($xy$) along the axial axis. The results are displayed in Fig. \ref{fig:mre gs with models}. We also compute the RSE over the full 3D calculation volume (see $\text{RSE}_\text{3D}$ in the labels in each panel of Fig. \ref{fig:mre gs with models}). 

It is shown that the SP-CZT* PSF has the least RSE relative to the RW-GS with a 3D RSE value of $0.0019 10^{-3}$. This is followed by the RW, Sinc-R*, F-Shell* then SP-FFT*. The 2D RSE in those Fourier-based models become higher at higher depth due to possible Fourier wrap-around present in the PSFs. The vector PSFs, vPSF and PSFGEn, have higher RSE values than the Fourier-based models with the RW-GS. The error in those models are more significant near the focus due to the violation of the missing cone. This concept of missing cone is discussed in details in Section \ref{comparison:missing cone}. The highest RSE are in the scalar PSFs and the errors stand at $4.8644 10^{-3}$ and $3.6971 10^{-3}$ for the GL and sPSF respectively. These reflect the limitation of the validity of the scalar approximation for a system with high numerical aperture.

\subsection{Violation of the missing cone}\label{comparison:missing cone}
\begin{figure}[htbp]
	\centering
	\includegraphics[width= 0.25\textwidth]{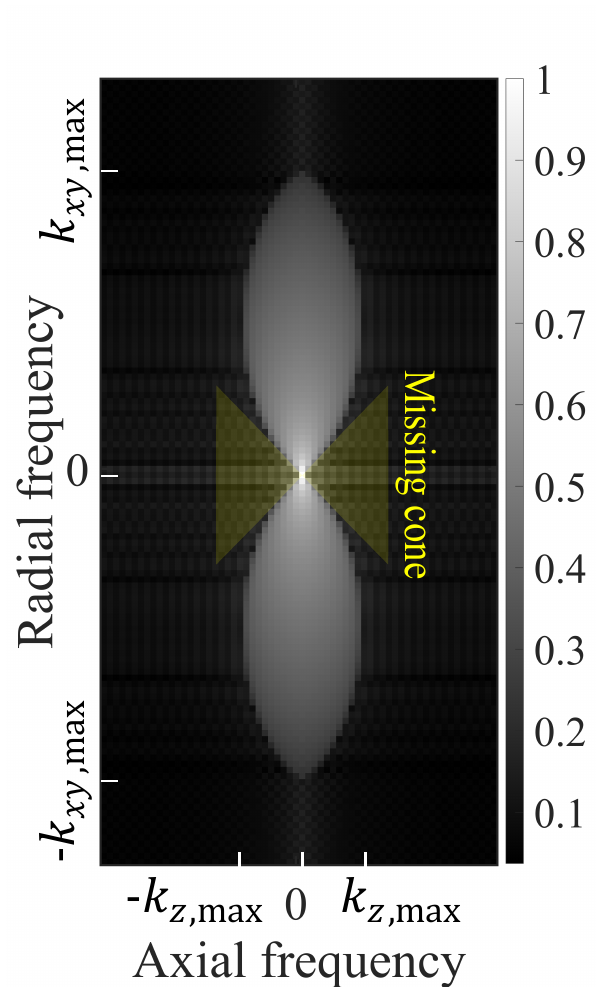} 
	\includegraphics[width= 0.48\textwidth]{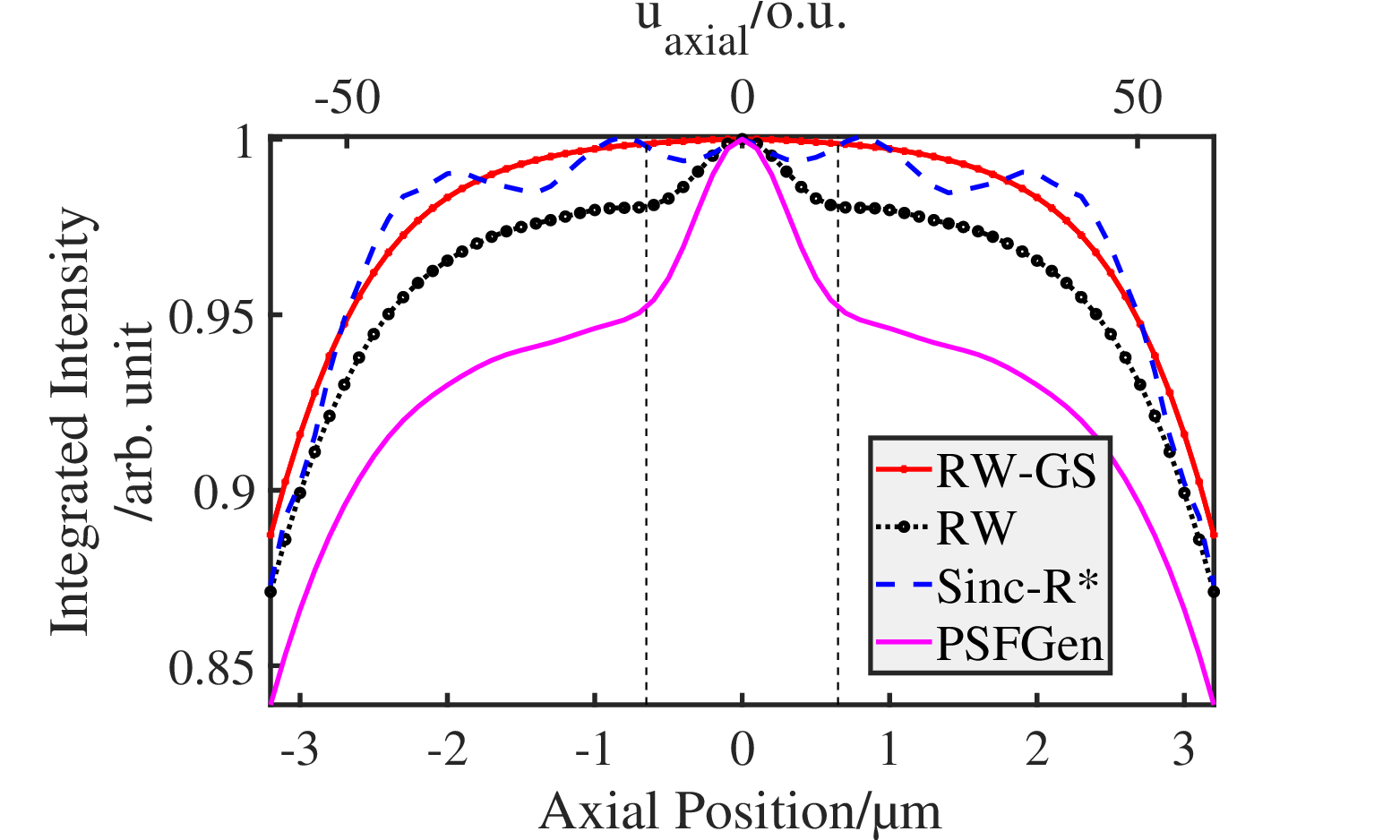} 
	\begin{minipage}{0.3\textwidth}
		\subcaption{\label{fig:otf-missingcone} }
	\end{minipage}
	\begin{minipage}{0.48\textwidth}
		\subcaption{\label{fig:II RW-SR-PSFGen} }
	\end{minipage}

	\includegraphics[width= 0.48\textwidth]{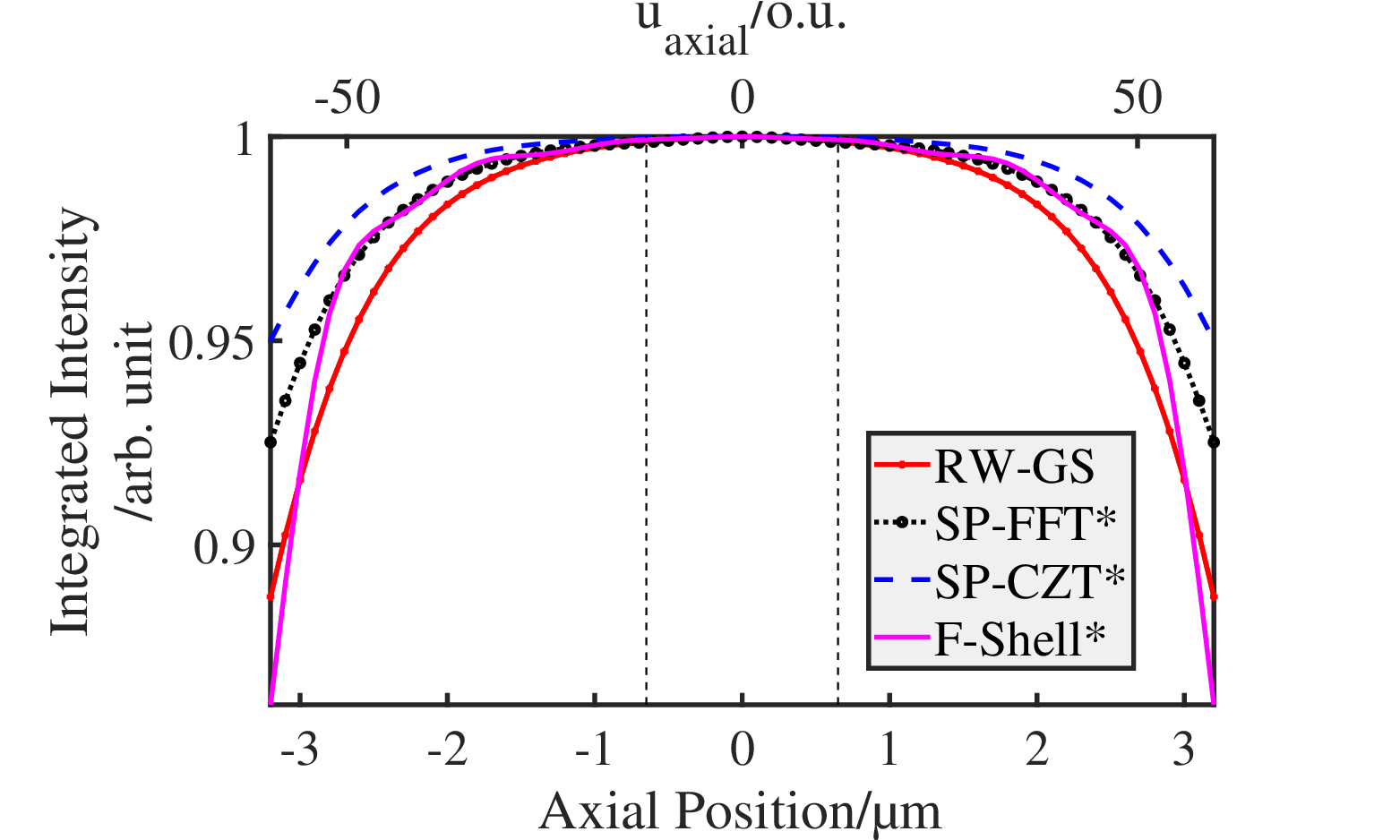}
	\includegraphics[width= 0.48\textwidth]{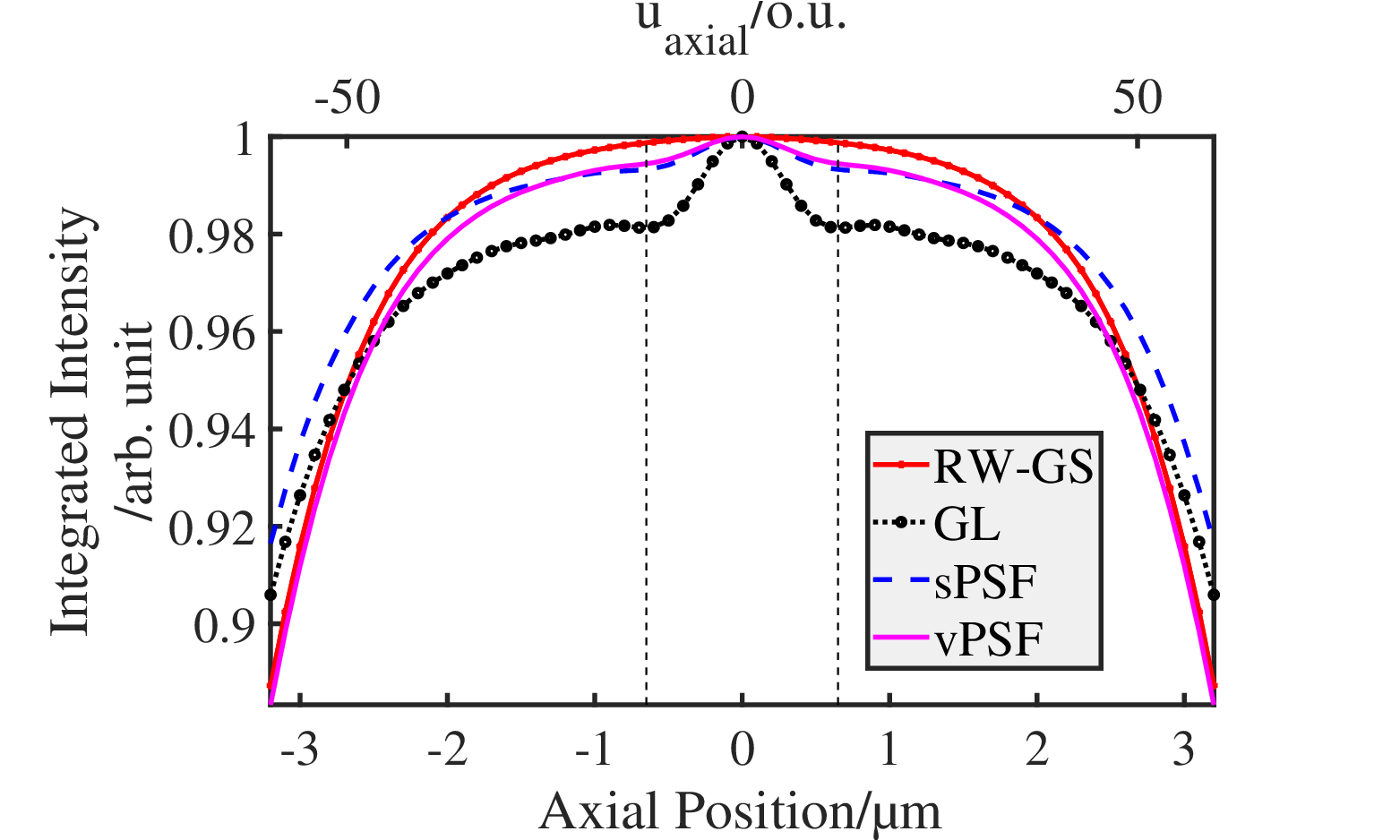}
	
	\begin{minipage}{0.48\textwidth}
		\subcaption{\label{fig:II RW-SP-CZT-VS} }
	\end{minipage}
	\begin{minipage}{0.48\textwidth}
		\subcaption{\label{fig:II RW-GL-sPSF-vPSF} }
	\end{minipage}
	
	\caption{(a) Representation of the optical transfer function (OTF) calculated using RW-GS and displayed at gamma of $0.2$. The missing cone is presented in yellow. (b, c, d) Observation of the violation of the missing cone by calculating and presenting the integrated intensity along $z$ of the PSFs. The profile which corresponds to RW-GS is displayed in all figures (b), (c), and (d) in red solid line as a reference.}
	\label{fig:integrated intensity}
\end{figure}

In wide-field microscopy, the frequency spectrum along the $k_z$-axis is missing apart from the $k_z=0$ position due to energy conservation in the system. Mostly independent of the focus position, all power emitted into angles of acceptance, as defined by the pupil, reaches the detector (if sufficiently large). This effect corresponds to a missing cone in the optical transfer function (OTF) that prevents the transmission of information about the object within this cone region (yellow cone in Fig. \ref{fig:otf-missingcone}). Out-of-focus light is distributed to different regions but, the power is still conserved as long as it resides somewhere on the detector. The integrated intensity at each $z$ position should therefore remain constant. This should be the case for any wide-field PSF as long as the PSF remains confined well within the calculation grid. Monitoring the integrated intensity for each $xy-$plane at each $z-$position to check for a violation of the missing cone, can, especially near the focus, be conclusive to reveal interpolation errors caused by mapping radial results to the rectilinear grid. In the calculation of the vector field in a finite grid, interpolation errors caused by insufficient data points for the computation can lead to the violation of the missing cone in the corresponding transfer function of the system. 

\begin{figure}[htbp]
	\centering
	\includegraphics[width= 0.4\textwidth]{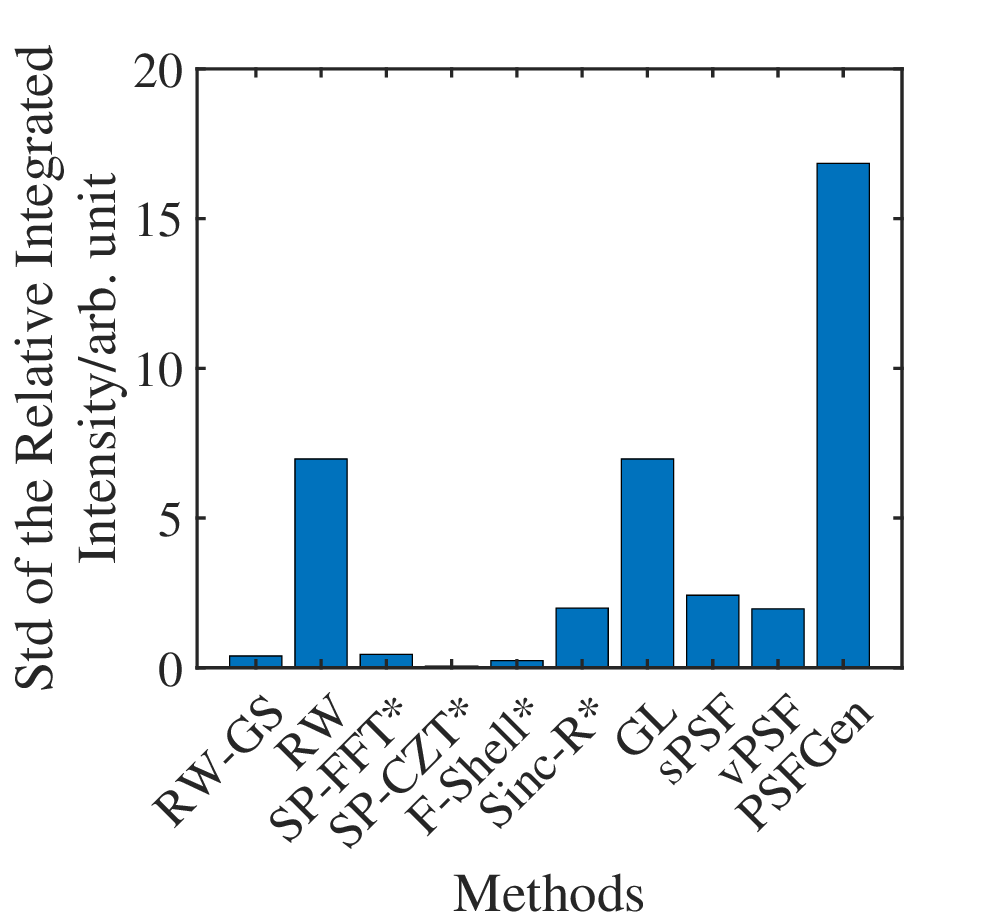} 
	\caption{Standard deviation of the integrated intensity over the \SI{20}{\percent} around the focus of the given $z-$range.}
	\label{fig:Std 20percent}
\end{figure}

A more precise observation of the violation of the missing cone can be made by zooming on the $z-$range around the focus and displaying the integrated intensity of each slice along the axial position $z$ (see Fig. \ref{fig:Std 20percent}).  To quantify this effect, the standard deviation of the integrated intensities within \SI{20}{\percent} around the focus of the given $z-$range (Std) is calculated and is plotted in Fig. \ref{fig:Std 20percent}. This \SI{20}{\percent} region is delimited by the two dashed vertical lines in Fig. \ref{fig:II RW-SR-PSFGen}, \ref{fig:II RW-SP-CZT-VS}, and \ref{fig:II RW-GL-sPSF-vPSF}.  The Std measures the non-uniformity of the laterally integrated intensity over the axial position range.

The Std , which corresponds to the SP-FFT* and the F-Shell* methods are very close to the Std of the highly sampled RW-GS while the SP-CZT* demonstrates the lowest value of Std. The methods described in this work, SP-FFT*, SP-CZT*, F-Shell* and Sinc-R* all have relatively low Std compared to the state-of-the-art PSFs: GL, sPSF, vPSF, and PSFGen. PSFGen is in the first position in violating the missing cone condition followed by RW computed with a standard sampling and GL. In 3D deconvolution routines applied to widefield data, preserving the missing cone property is paramount. 

\subsection{FFT wrap-around effect}\label{comparison:wrap-around}

To quantify the wrap-around effect, we calculate densely sampled PSFs using each technique on a large calculation window, denoted by $W_{\text{ref}}$ using the formula in Eq. \eqref{Eq:desired window without wrap-around Nx}. This PSF is denoted by $h_{\text{ref}}$. 

\begin{equation}
	W_{\text{ref}|x} = 2\times2\times \left( \frac{z_\text{max} \frac{k_{x|\text{max}}}{k_{z|\text{max}}} + 1.3 d_{x|\text{lim}}}{p_x}\right), 
	\label{Eq:desired window without wrap-around Nx}
\end{equation}
with $k_{x|\text{max}}$ and $k_{z|\text{max}}$ are the maximal radial and axial frequency components, $d_{x|\text{lim}}$ being the resolution limit, $p_x$ the pixel pitch and 1.3 is a heuristic factor. The first factor 2 in the expression of $W_{\text{ref}|x}$ is considered to double the half window and the second factor 2 is to sample the frequency space two times higher. The variables $z_\text{max}, p_x$ and $d_{x|\text{lim}}$ have the same units and $W_{\text{ref}|x}$ is expressed in pixels. The same amount of padding is also applied along $y-$axis.

Substantial wrap-around effects should be avoided within this large computation grid $W_{\text{ref}}$. In order to quantify the wrap-around effect of each PSF model, we then compute the PSF $h_W$ with the (typically smaller) window $W$ and compare it to $h_{\text{ref}}$ calculated with the window $W_{\text{ref}}$. The wrap-around effect in $h_W$ in relation to $h_{\text{ref}}$ is calculated by subtracting $h_W$ and $h_{\text{ref}}$ over the smaller of the two windows (see Eq. \eqref{Eq:epsilon error}) to obtain the amount of standing waves expressed in intensity values in the PSFs. No wrap-around effect is expected for $W>W_{\text{ref}}$. However, since the reference window may already show minor wrap-around effects, the error $\epsilon$ is expected to converge to a small non-zero constant for $W>W_{\text{ref}}$. 

\begin{figure}[htbp]
	\centering
	\includegraphics[width= 0.7\textwidth]{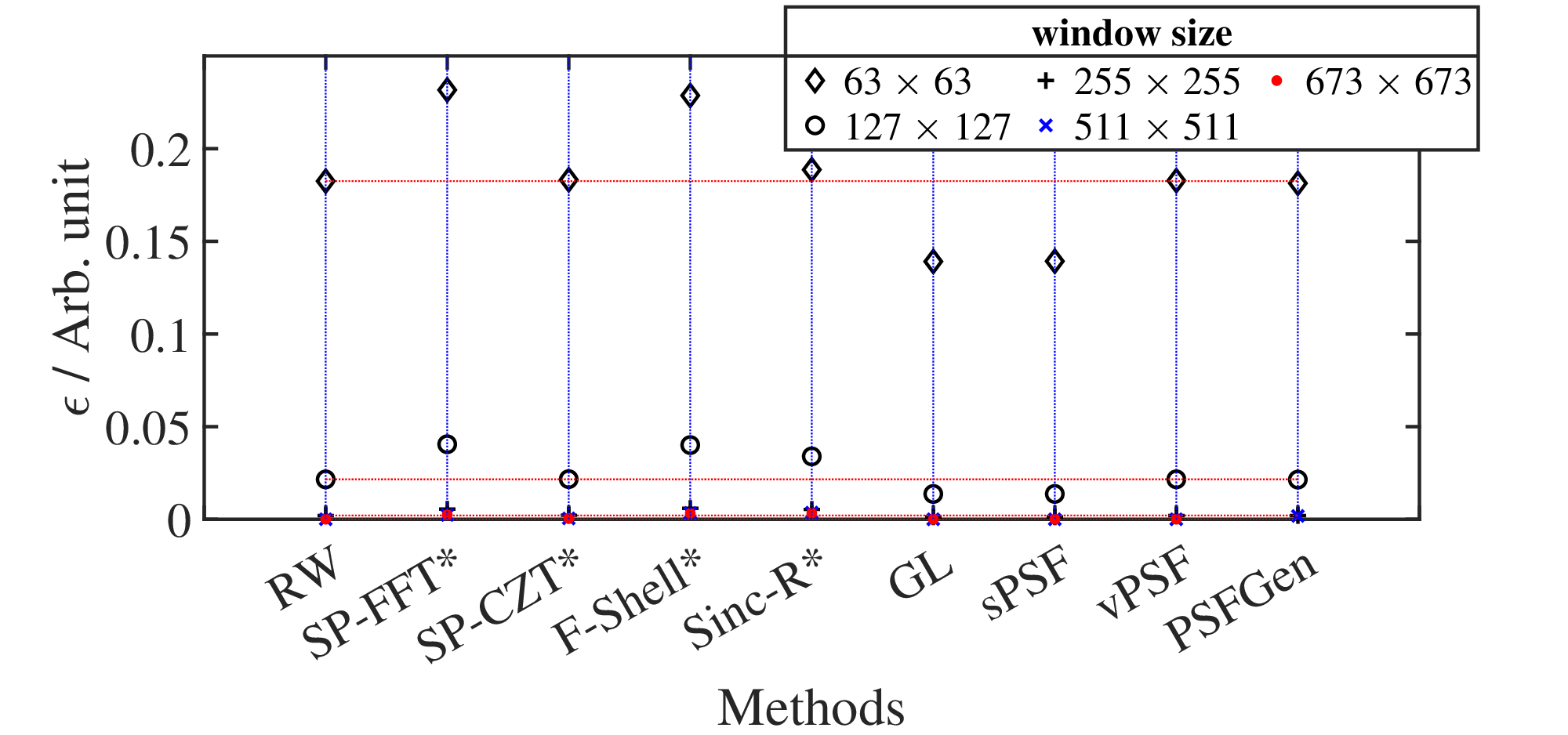} 
	
	\caption{Quantification of the wrap-around effect in dependence of the size of the calculation window ($x$ and $y$). The red dashed horizontal lines pass by the error $\epsilon$ of the RW for each given window size. The reference window is calculated to be $337\times 337$.}
	\label{fig:wrap-around}
\end{figure}

The $W_{\text{ref}}$ is calculated to be $337 \times 337$ in $x$ and $y$. We compute different window size: $63 \times 63$, $127 \times 127$, $257 \times 257$, $511 \times 511$, and $673 \times 673$. The $z-$range is 65 pixels (\textit{i.e.} 65 $xy-$slices along $z$). The rest of the parameters are the same as previously described in the introduction of Section \ref{Sec:comparison methods}. GL, sPSF, vPSF, PSFGen, which are state-of-the art methods and RW do not employ any FT operators and are therefore not prone to wrap-around problems. We observe from Fig. \ref{fig:wrap-around} that the error $\epsilon$ of the scalar state-of-the-art GL and sPSF PSFs fall far below the red dashed lines passing at the RSE of RW. The RSE of the SP-CZT* falls on the line passing horizontally on RW, vPSF and PSFGen for all the different windows. The Sinc-R* presents an error $\epsilon$ slightly above the red lines while SP-FFT* and F-Shell* fall at about 0.05 in RSE above the red lines. This means the wrap-around effect in the SP-CZT* has been completely suppressed while there is still some remaining wrap-around problem in the SP-FFT* and F-Shell* methods. The error in the value of $\epsilon$ is less than the errors in SP-FFT* and F-Shell* for the Sinc-R*. This is still an error in the Sinc-R* method that should be improved in the future. All the $\epsilon$ errors converge to a constant as expected for window bigger than the reference window $327 \times 327$.

\subsection{Computational time}\label{comparison:computation time}

\begin{figure}[htbp]
	\centering
	\includegraphics[width= 0.7\textwidth]{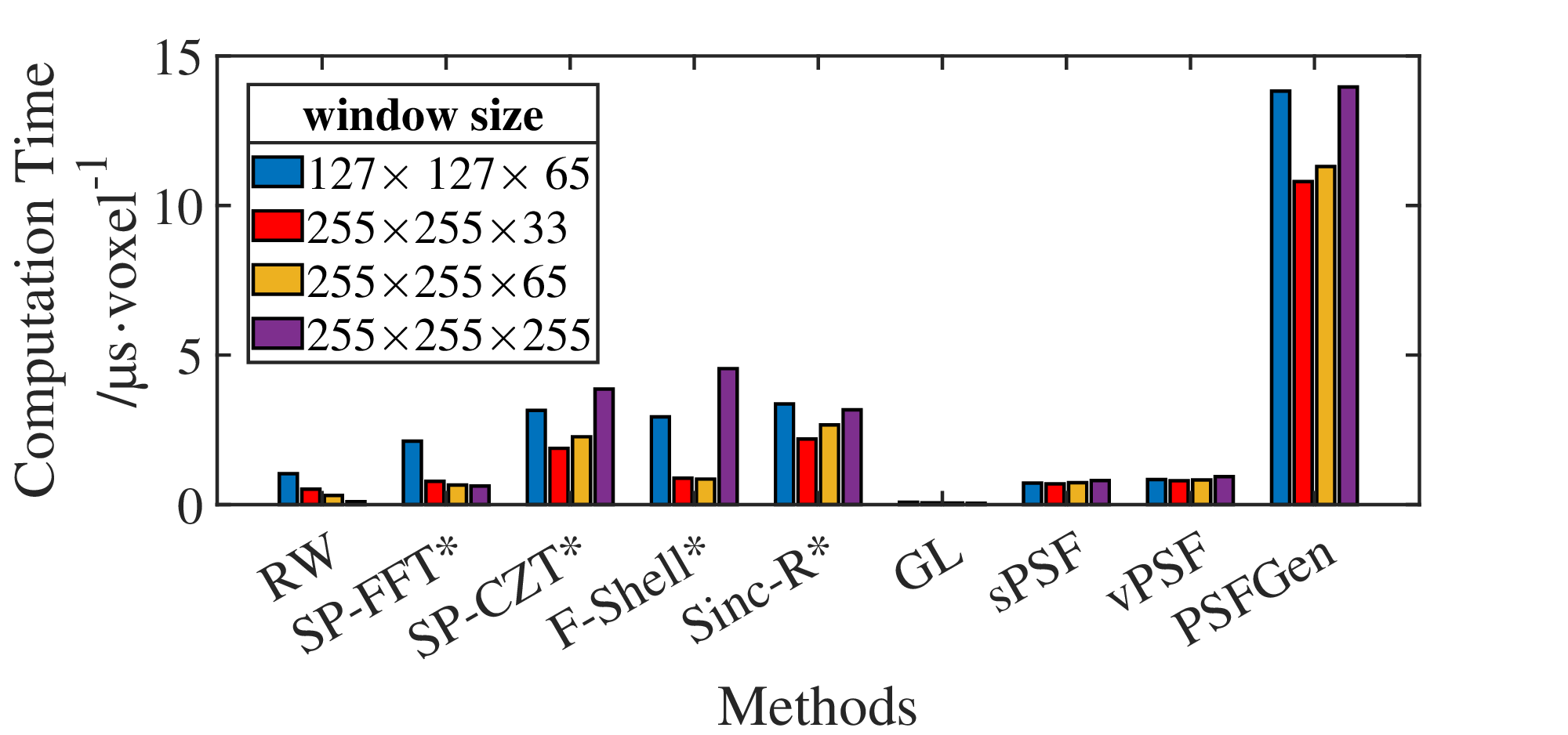} 
	\caption[Computational time of each technique per voxel in \SI{}{\micro\second}]{Computational time of each technique per voxel in \SI{}{\micro\second}.}
	\label{fig:computation time}
\end{figure}

The estimation of the speed of the various algorithms in this section was performed using Windows 10, 64-bit, 
Intel\textsuperscript{\tiny\textregistered} Core\textsuperscript{\tiny TM} i5-3570S CPU @ 3.10GHz, 8,0GB RAM, Intel HD Graphics. The profiles of the computation time per voxel for each technique at four different window sizes are displayed in Fig. \ref{fig:computation time}.

In the RW model, the three integrals of the electric field \cite{RW2} are evaluated numerically by exploiting a cylindrical coordinate system for the integration. This numerical integration is computationally advantageous since only the field in the centred radial axis versus the axial position $z$ is calculated. RW, GL, sPSF and vPSF exploit the radial symmetry in their computation. This has the disadvantage that non-spherical aberrations cannot easily be represented. The accuracy of the RW model is a function of the number of data points used in the numerical integration.

3D SP-FFT* and SP-CZT* compute PSFs slice by slice for each $z$ position. Using a code profiler under Matlab, we observed that about 70$\%$ of the computational time of the SP-FFT* technique is dominated by the FFT operation. The algorithmic complexity of each 2D slice in the SP-FFT* is given by $O(N_xN_y\log(N_xN_y))$. The computational time per voxel for different window sizes therefore remains constant, while the total computation time of the volume PSF scales linearly with the total number of pixels. 

Reflecting on Eq. \eqref{Eq:czt formula}, the CZT operation requires 3 FFTs to compute. In the SP-CZT* method, a new and bigger window is calculated such that the field at higher depth would not suffer from wrap-around effects. The run-time depends on both, lateral size and $z-$depth. This explains the relative higher computational time per voxel in SP-CZT* compared to SP-FFT*.

The Fourier shell method (F-Shell*) computationally depends on the interpolation of the coefficients of $k_z$ at different sub-pixels. By analysing the algorithm with the code profiler under MATLAB, this interpolation process takes up about $70\%$ of the total computation cost. If the $z-$range becomes significantly larger, a large number of coefficients also needs to be interpolated. The computation time therefore mainly depends on the number of iterations and the kernel size. Those coefficients are stored on the disk of the computer and can be accessed easily without any heavy computation, reducing the computation time of a second run by up to 20$\times$.

In the computation of the Sinc-R* PSF, the Fourier wrap around in the calculation of wave propagation is avoided by starting from the real-space ($\sinc(r)$) representation of a Fourier-shell and limiting and modifying it further in Fourier space. Artefacts that may be present in the calculation from the edge of the grid during this 3D FFT. This can be avoided by damping the edge or extending the window by 25$\%$. The difference in the computation with or without this 25$\%$ extension is not significant. It is advisable to consider it in the computation as more information can be preserved by it. It should be noted that modifications in Fourier-space lead to convolutions in real space, which can also cause wrap-around artefacts, but these seem to be relatively minor.
In the Sinc-R method, the required $z-$range needs to initially be highly sampled leading to an heavy computation, which is not required for the other propagation methods.

The state-of-the-art GL method discussed here is implemented using a Bessel series approximation of the GL method \cite{li2017fast} over a single $xz$ slice. A radial asymmetry in the PSF is assumed and a piece-wise linear interpolation is used to compute the whole 3D volume PSF on a rectilinear grid. The model's computational time and accuracy are inversely dependent. Both depend on the number of basis in the Bessel series and the sampling number. The computation time of the PSF corresponds to the default number basis vectors, which is equal to 100 Bessel functions and number of coefficient parameters equal to 1000. These make the GL PSF calculation the fastest, albeit not very precise scalar method.

sPSF, vPSF and RW compute the numerical integration of the electric field in the image plane using Simpson's rule \cite{aguet2009super}. The methods are relatively fast and the computational time per voxel seems to be independent of the $z-$range. Lastly, the software PSFGen is the most expensive in time and in memory among the PSF models, which we compare to in this work. 


\section{Conclusion}\label{conclusion}

In this work, we presented a general approach for calculating the $3$D PSF of a system satisfying the Abbe sine condition using Fourier based techniques. The PSFs models, with other state-of-the-art PSF calculation methods, are compared to a defined gold standard, which consists of an highly sampled Richard and Wolf model. We account for the digital Fourier transform pitfalls in the calculation and present different strategies to avoid them. The SP-CZT* model is proven to be the most accurate among our models. This accuracy comes at a cost in terms of its computation particularly at large field depths. The SP-FFT* and F-Shell* have similar accuracy and behaviour. In their computation, standing waves caused by wrap-around (which can also be interpreted as undersampling of the phase near the edge of the pupil) in the PSF are not fully avoided when the calculation grid is insufficient for the given depth of field but reduced. These two methods are however accurate and fast enough when a large depth of field is not required. The Sinc-R* model presents an attractive feature for easy implementation of a volume PSF largely avoiding the wrap-around problem. Overall the methods described in this work satisfy the physical condition of the imaging better than the state-of-the-art PSFs. They are already computationally efficient compared to the state-of-the-art given the fact that there is no use of radial symmetry to speed up their calculation. Our PSFs models can accommodate any non-circular aberration and can easily be modified to account for a tilted plane such as a tilt in a coverslip for instance. The models are also reproducible and can be easily extended to represent different imaging modalities. An experimental validation of the PSF models presented in this work is essential to verify the experimental accuracy of these models.

\begin{backmatter}
\bmsection{Funding}
This work was funded by the DAAD through the African Institute for Mathematical Sciences and Stellenbosch University, and Friedrich Schiller University Jena. This work was also supported by the German Research Foundation (DFG) through the Collaborative Research Center PolyTarget 1278, project number 316213987, subproject C04 and the Council for Scientific and Industrial Research (CSIR), project number LREQA03.

\bmsection{Acknowledgments}
The authors wish to acknowledge Herbert Gross and Norman Girma Worku for the CZT function, Peter Verveer for the first version of the DIPimage Library of the RW code, Colin Sheppard for valuable discussions and, the nano-imaging research group especially Ren\'e Lachmann, Ronny F\"orster and Jan Becker at the Leibniz Institute of Photonic Technology, Jena, Germany for their contributions to this work. 

\bmsection{Disclosures}
The authors declare no conflicts of interest.

\bmsection{Code Availability}
The MATLAB toolbox for Fourier-based PSF modeling is publicly available on GitHub \cite{PSFToolbox}

\end{backmatter}

\bibliography{sample}

\end{document}